\documentclass[manuscript,nonacm]{acmart}

\setcopyright{none}
\renewcommand\footnotetextcopyrightpermission[1]{}

\AtBeginDocument{%
  }

\acmISBN{978-1-4503-XXXX-X/2018/06}
\usepackage{booktabs}
\usepackage{tabularx}
\usepackage{progressbar}
\usepackage{tikz}
\usepackage{graphicx}
\usepackage{subcaption}

\usepackage{fontawesome5}

\newcommand{\primary}{\faCircle}
\newcommand{\secondary}{\faCircle[regular]}
\newcommand{\integration}{\faAdjust}

\usepackage{graphicx}
\usepackage{subcaption}
\usepackage{xcolor}
\usepackage{tabularx}
\usepackage{enumitem}
\usepackage{array}

\definecolor{cKnowledge}{HTML}{1F6F63}
\definecolor{cSynthesis}{HTML}{6A4794}
\definecolor{cMethods}{HTML}{A37A0B}
\definecolor{cFraming}{HTML}{4171BF}
\definecolor{cIsolation}{HTML}{BA2F2F}
\definecolor{cNovelty}{HTML}{D13B7D}

\newlist{finditems}{enumerate}{1}
\setlist[finditems]{label=\Roman*., leftmargin=1.4em, itemsep=3pt, topsep=2pt, parsep=0pt, font=\bfseries}

\begin{document}

\title{Orchestrating GenAI for Interdisciplinary Research}

\author{Shirley Anugrah Hayati}
\email{hayat023@umn.edu}
\affiliation{%
  \institution{University of Minnesota}
  \city{Minneapolis}
  \state{Minnesota}
  \country{USA}
}

\author{Moyan Zhou}
\email{zhou0972@umn.edu}
\affiliation{%
  \institution{University of Minnesota}
  \city{Minneapolis}
  \state{Minnesota}
  \country{USA}
}

\author{Patricia Anugrah Setiani}
\affiliation{%
  \institution{University of Minnesota}
  \city{Minneapolis}
  \state{Minnesota}
  \country{USA}
}

\author{Ruizi Wang}
\affiliation{%
  \institution{University of Minnesota}
  \city{Minneapolis}
  \state{Minnesota}
  \country{USA}
}

\author{Joseph Chee Chang}
\affiliation{%
  \institution{Allen AI Institute}
  \city{Seattle}
  \state{Washington}
  \country{USA}
}

\author{Dongyeop Kang}
\email{dongyeop@umn.edu}
\affiliation{%
  \institution{University of Minnesota}
  \city{Minneapolis}
  \state{Minnesota}
  \country{USA}
}



\begin{abstract}
As researchers tackle interdisciplinary problems, they face the need to deepen expertise in primary areas while rapidly acquiring knowledge in secondary domains. Generative AI (GenAI) is increasingly positioned to meet this need, from general-purpose chat assistants to Deep Research tools marketed as autonomous research agents. Prior work has examined how researchers use GenAI to support single-discipline or general research tasks. However, we know little about the goals and GenAI practices in interdisciplinary research. We conducted a longitudinal study and semi-structured interviews with 15 interdisciplinary researchers to examine how interdisciplinary researchers actually orchestrate GenAI. Findings show that researchers leaned on GenAI to fill knowledge gaps while maintaining epistemic agency for novelty discovery. We also uncovered an expertise paradox: GenAI outputs were hardest to verify when most needed. Our empirical insights motivate GenAI designs that calibrate verification to researchers' expertise, nudge toward cross-domain synthesis, and adapt prompting and outputs to disciplinary conventions.
\end{abstract}



\received{20 February 2007}
\received[revised]{12 March 2009}
\received[accepted]{5 June 2009}

\maketitle

\fancyhf{} 
\fancyfoot[C]{\thepage} 
\renewcommand{\headrulewidth}{0pt}
\renewcommand{\footrulewidth}{0pt}

\section{Introduction}
Interdisciplinary research is recognized as essential for advancing scientific knowledge and addressing today's most pressing challenges \cite{rana2025interdisciplinary, rong202540, repko2020interdisciplinary}. Real-world problems often span multiple disciplinary boundaries and require knowledge, methods, and perspectives that cannot be adequately addressed within a single domain \cite{van2015interdisciplinary, Zheng2024DiscipLinkUI}. This matter is highly relevant to human-computer interaction (HCI) as an inherently interdisciplinary field that draws on theories and practices from various disciplines such as computer science, psychology, design and social sciences \cite{mackay1997hci, carroll2003hci, oppenlaender2025past, kaltenhauser2025thechurns, chen2024hci}. 
As a result, interdisciplinary research asks scholars to be credible in domains they have limited training, such as integrating computational methods to a virology problem as a machine learning researcher or looking into labor-economics literature as an operations management scholar. Unlike learning within one's primary discipline, where researchers can draw on established communities and institutional resources, interdisciplinary researchers face a unique challenge of independently navigating unfamiliar knowledge across fields \cite{bates1996learning, newby2011entering, spanner2001border, dalton2022interdisciplinary, bromme2000beyond}. In practice, they must balance advancing their primary discipline with keeping pace with developments in secondary areas. Collaborating with domain experts is one way to bridge this gap, but access to such expertise is not equally available in research communities. In inherently interdisciplinary fields such as HCI, researchers may have more opportunities to work with domain expert collaborators from different disciplines. Researchers situated in more disciplinary-specific communities, however, may have fewer opportunities to find relevant collaborators when their work extends into another field, such as a social scientist incorporating natural language processing methods into their research. Resource and budget constraints can further limit such collaborations, so researchers must independently navigate unfamiliar domains \cite{brown2015interdisciplinarity}.

As GenAI continues to be integrated into the research life cycle \citep{zhang2025exploring, luo2025llm4sr, ren2025towards}, prior work in HCI has examined GenAI use in various research stages. For example, the use of GenAI in early stages of the research workflow, including research ideation, novel hypothesis generation, and literature review, has been studied ~\cite{ pu2025ideasynth, liu2024ai, lin2025seeking, radensky2024scideator, wang2024evaluating}. In the later stage of research, such as writing, GenAI can generate surface-level text and also higher-level cognitive processes~\cite{gero2022sparks, lee2022coauthor, monge2025investigating, dhillon2024shaping}. This trend reflects the expanding role of GenAI in reshaping knowledge-driven research practice in various domains, including qualitative studies, social sciences ~\cite{pang2025understanding, schroeder2025large, gao2024collabcoder, gebreegziabher2023patat, grossmann2023ai, zheng2025evalignux}, and HCI itself \cite{pang2025understanding}. However, these studies have mostly focused on general research contexts or research within a single discipline. Meanwhile, interdisciplinary researchers face distinct needs: acquiring knowledge in less familiar domains and integrating insights from multiple disciplines \cite{reddy2025towards, morris2023scientists, fecher2025friend, eger2025transforming, liao2024llms, ye2024language}. Addressing this gap is timely and important as interdisciplinary research continues to grow and agentic AI tools are increasingly marketed as research assistants, but how these capabilities support interdisciplinary research in practice remains less understood.

This question sits at the center of HCI's concern with human–AI interaction as interdisciplinary researchers may move between areas where they have different levels of expertise, even within the same research goal. Our work extends previous HCI research that has examined the use of LLMs for supporting knowledge sense-making in research workflows, but we focus on the needs of interdisciplinary researchers ~\cite{liu2024selenite, drosos2024s, yun2025generative, kittur2013costs, kittur2014standing}. 

Concretely, we aim to answer the following research questions: \textbf{How do interdisciplinary researchers orchestrate GenAI tools to navigate challenges in their research?}

Through a study of 15 interdisciplinary researchers from diverse backgrounds (e.g., nutrition sciences, engineering, education, sociology), we investigate how they orchestrate two GenAI tools in their everyday research practice: (1) the standard GenAI where the user prompts the model and can iteratively steer its response, and (2) Deep Research \cite{chatgptdeepresearch2025} which is specifically tailored to assist researchers as it produces massive search and retrieval when providing the answers. By giving participants access to both resources, we examine when and why they turn to each one as their research needs evolve. Participants used both tools in their real-world research workflows over 1-2 weeks. We triangulated longitudinal tracking documents, interaction logs, and post-interaction interviews to investigate their orchestration strategies.

Findings show that participants mainly used GenAI for breadth-oriented exploration rather than depth-oriented reasoning, while retaining responsibility for filtering and integration. Moreover, though GenAI was valuable for entering secondary fields, interdisciplinary researchers were less equipped to verify its outputs. On the other hand, in their primary domains, where researchers had deeper expertise, they scrutinized GenAI outputs more critically, echoing prior findings on how expertise shapes algorithm aversion and appreciation \cite{hou2021expert, chen2025ai}. This asymmetry raises the concern that novices, who are less equipped to catch hallucinated or unsupported claims, may also be the most willing to accept them.

Through this work, we make the following contributions:
\begin{itemize}
\item \textbf{Empirical findings of interdisciplinary research needs and how researchers orchestrate different GenAI tools.} We identify six challenges that motivate interdisciplinary researchers to seek GenAI support and show how these needs shape GenAI resource allocation. Researchers relied on GenAI for knowledge discovery and technical assistance in their secondary research areas. In contrast, challenges related to their primary expertise, such as discovering novelty, were harder to delegate to GenAI.

\item \textbf{An expertise paradox in AI-assisted interdisciplinary research.} Researchers valued GenAI in their secondary areas for quickly surfacing unfamiliar terminology and providing background context. In primary areas, deeper expertise led them to scrutinize GenAI limitations more critically. This suggests that GenAI's perceived value does not merely depend on the output quality but also on researchers' existing knowledge. We term this asymmetry as an ``expertise paradox.''

\item \textbf{Design implications for supporting interdisciplinary knowledge construction.} GenAI was effective at broadening researchers' access to knowledge across domains, but meaningful interdisciplinary integration still depends on researchers to frame and make the connections themselves. We translate these findings into concrete design opportunities, including verification support via language style translation in less familiar domains, nudging researchers toward cross-disciplinary synthesis, adapting output formats to disciplinary practices, and providing prompting guidance for less familiar domains.

\end{itemize}

\section{Related Work}
\subsection{GenAI as Research Assistants}
Researchers have developed GenAI tools for different stages of research workflows, including idea generation \citep{gero2022sparks, pu2025ideasynth, radensky2024scideator}, literature review \citep{wang2024evaluating, choe2024supporting, lee2024paperweaver, agarwal2024litllms, tang2024large, zhou2025hypothesis}, and manuscript writing \citep{radensky2024posts, zhou2025hypothesis, heidt2025ai, monge2025investigating, zhang2025paperbridge}. For example, Ideasynth \citep{pu2025ideasynth} helps researchers develop initial ideas into literature-grounded summaries so they are able to explore alternative ideas. LitLLMs \citep{agarwal2024litllms} decomposes literature review writing into retrieval and planning for structured drafts. Radensky et al. \citep{radensky2024posts} designed a writing assistant for planning, drafting, and revising while Zhang et al. \citep{zhang2025paperbridge} supports exploration of diverse perspectives for research narratives. These examples suggest that GenAI is not replacing researchers' judgment but redistributes human cognitive effort toward evaluation, integration, and stewardship of its outputs \cite{lee2025impact, woodruff2024knowledge, venkit2024searchenginesaiera, messeri2024artificial}. 


Despite these advances, hallucinations \citep{tang2024large}, plagiarism \citep{gupta-pruthi-2025-glitters}, and illusions of understanding \citep{messeri2024artificial} remain barriers to GenAI use in research. Recent studies also document gaps between promising GenAI-generated ideas and their execution \citep{sican,si2025ideation}. Commercial Deep Research systems \cite{gemini2025, perplexity2025, chatgptdeepresearch2025} represent a move toward more assistance for researchers compared to the default chat modes (standard GenAI). The standard mode typically performs no or limited retrieval-augmented generation (RAG), but Deep Research conducts multiple steps of reasoning and retrieval to produce more insightful and longer reports with citations.


However, systems specifically designed for interdisciplinary work remain sparse \cite{Liu2024PersonaFlowDL,Zheng2024DiscipLinkUI,guo2024personalized,bao2025words}. Existing systems mainly support terminology translation across fields or interdisciplinary ideation, but we do not know how GenAI supports the broader process of navigating and integrating knowledge across disciplines. We address this gap by empirically examining how interdisciplinary researchers engage with GenAI in diverse stages of their natural research workflows.

Survey and qualitative studies further show that researchers are already incorporating LLMs into everyday research practices. Liao et al. \citep{liao2024llms} found that 81\% of 816 published researchers reported using LLMs for tasks including information seeking, editing, and ideation. Morris et al. \citep{morris2023scientists} similarly found scientists willing to augment writing, coding, and literature review with LLMs while remaining reluctant toward full automation. Qualitative work shows that researchers strategically negotiate LLM use according to their goals and disciplinary norms \citep{monge2025investigating}. Researchers may frame LLMs as collaborators rather than passive tools \citep{zhang2025collab, jin2025high}, while also expressing concerns about disclosure, authorship, bias, and privacy \citep{schroeder2025large, kapania2025m, zhang2025secret}. Such concerns, together with risks of illusory productivity \citep{messeri2024artificial}, help explain why researchers often treat LLMs as augmentative assistants under human oversight rather than autonomous agents \citep{ren2025towards, zhang2025collab}.

Building on this work, our work investigates how interdisciplinary researchers orchestrate GenAI tools to address research challenges such as knowledge gaps. Specifically, we study what \textit{interactions} researchers perform to achieve what \textit{goals} in which \textit{disciplines}, providing insight into the design of GenAI research assistants for interdisciplinary research.


\subsection{Interdisciplinary Research}
Interdisciplinary research involves researchers from two or more disciplines working toward a shared goal by integrating perspectives, methods, and insights across fields \citep{aboelela2007defining, daniel2022challenges, repko2020interdisciplinary, vladova2025whom}. It is increasingly important for complex problems that cannot be addressed within a single discipline \citep{aboelela2007defining, barkovic2010challenges, newman2024promoting}. Diverse expertise can enable more holistic and creative outcomes \citep{klein2008evaluation, siedlok2014organization, purvis2023critical}, while also fostering innovation, exploration, collaboration, and individual benefits such as curiosity and creativity \citep{newman2024promoting, vantard2023interdisciplinary, spence2024and, lamain2024finding}.

Yet interdisciplinary research is difficult to enact. At the disciplinary level, epistemological and methodological differences can create tension and communication difficulties \citep{aboelela2007defining, siedlok2014organization}. At the collaborative level, inconsistent terminology, differences in valued problem spaces, and limited belonging for researchers without a ``home'' department can hinder teamwork \citep{teare2020extending, daniel2022challenges, purvis2023critical}. Institutional barriers include physical separation, bureaucracy, misaligned incentives, limited funding, short-term evaluation pressures, and scarce publication venues \citep{vladova2025whom, newman2024promoting, sun2021interdisciplinary, vantard2023interdisciplinary}. Addressing these challenges requires researchers to reconcile disciplinary tensions and negotiate shared goals \citep{aboelela2007defining, mccance2024measuring, klein2008evaluation}. Prior work identifies strategies such as cross-disciplinary communication, digital platforms for finding collaborators, and shifting team perspectives toward collective goals \citep{bridle2013preparing, sharanowski2025social, mather2024rowing}.

The growth of interdisciplinary research has also produced researchers who themselves maintain interests across fields \citep{sun2021interdisciplinary, lamain2024finding, purvis2023critical}. Such researchers must balance depth in a primary field with breadth across secondary areas \citep{smith2007role}. Discipline expertise entails deep knowledge of how a field's data are collected, analyzed, interpreted, and communicated \citep{daniel2022challenges}; interdisciplinary researchers therefore often develop a core area of expertise while gaining sufficient knowledge of other fields to connect perspectives and methods \citep{waldman2013interdisciplinary, kelly2019ten, barkovic2010challenges}.

HCI itself is fundamentally interdisciplinary. Blackwell \cite{blackwell2015hci} characterizes HCI as an inter-discipline and ``trading zone'' connecting engineering, design, and the human sciences, while Harrison et al. \cite{harrison2007three} trace HCI's development through paradigms rooted in different disciplinary traditions. Other HCI work examines problem-driven interdisciplinary collaborations, including HCI with health research \citep{singh2017hci, blandford2018seven, agapie2022using, agapie2024conducting}. While this work has documented motivations, benefits, and challenges of such collaborations, less is known about how interdisciplinary researchers themselves navigate competing disciplinary commitments and use tools to balance primary-field depth with secondary-field breadth.

We address this gap by examining how interdisciplinary researchers use LLMs to support these distinctive research needs, focusing on the practices through which they negotiate depth and breadth and how these practices shape human-AI collaboration.

\section{Methods}

\begin{table*}[]
    \centering
    \small
    \begin{tabularx}{\textwidth}{ @{}p{0.4cm} p{1.5cm} p{2.3cm} p{2.2cm} p{2.2cm} p{2.7cm} p{1.2cm}@{} }
    \toprule
    \textbf{ID} & \textbf{Academic} &\textbf{Field} & \textbf{Primary Research} & \textbf{Secondary } & \textbf{Research} & \textbf{GenAI}\\
    & & & & \textbf{Research} & \textbf{Stage} & \textbf{Familiarity}\\
    \midrule
    P1 & Junior PhD & Computer science & Mixed reality & Robotics  & Literature review & \progressbar{0.9} \\[0.7ex]
    P2 & Senior PhD & Nutrition sciences & Gut brain axis & Gut microbiome & Writing manuscript & \progressbar{0.7}\\[0.7ex]
    P3 & Junior PhD & Computer science & Visualization & Generative AI & Data processing & \progressbar{0.7}\\[0.7ex]
    P4 & Postdoc & Educational psychology & Learning science & Data science & Planning \& ideation & \progressbar{0.9}\\[0.7ex]
    P5 & Senior PhD & Computer science & Explainable machine learning & Virology & Planning \& ideation, literature review, data collection & \progressbar{0.9}\\[0.7ex]
    P6 & Junior PhD & Industrial \& systems engineering & Optimization & Machine learning & Planning \& ideation, literature review & \progressbar{0.9}
\\[0.7ex]
    P7 & Junior PhD & Chemistry & Chemistry & Biology & Data collection, experiment, analysis & \progressbar{0.3}\\[0.7ex]
    P8 & Junior PhD & Industrial engineering & Operations research & Mathematical programming & Writing manuscript & \progressbar{0.7}\\
    P9 & Senior PhD & Feminist studies & Transnational feminism & Humor studies & Literature review & \progressbar{0.7}\\[0.7ex]
    P10 & Senior PhD & Bioinformatics & Computer science & Human genetics  & Planning \& ideation, literature review, data collection, data processing & \progressbar{0.8}\\[0.7ex]
    P11 & Junior PhD & Sociology & Cultural sociology & Natural language processing & Planning \& ideation, data collection & \progressbar{0.6}\\[0.7ex]
    P12 & Senior PhD & Supply chain \& operations & Operations management & Labor economics & Paper revision & \progressbar{0.9} \\[0.7ex]
    P13 & Postdoc & History & American history & Memory studies & Writing manuscript & \progressbar{0.1}\\
    P14 & Senior PhD & Chemistry & Polymer chemistry & Materials science & Data collection, data processing, experiment, analysis & \progressbar{0.1}\\
    P15 & Junior PhD  & Chemistry & Organic chemistry & Cellular biology & Data collection, data processing, experiment, analysis, writing manuscript & \progressbar{0.1}\\
    \bottomrule
    \end{tabularx}
    \caption{An overview of 15 interdisciplinary researchers who participated in our study.}
    \label{tab:participant_details}
\end{table*}

\subsection{Study Design}

Our study consisted of three phases: (1) a pre-interaction survey, (2) a 1-2 week longitudinal interaction period, and (3) a one-hour semi-structured interview. During the interaction phase, participants used two GenAI modes, standard GenAI and Deep Research, and documented their research activities and newly acquired knowledge as a diary study \cite{bolger2003diary, hyers2018diary, rieman1993diary}. We recruited participants across research stages and interdisciplinary backgrounds to capture how researchers integrate LLMs into their naturalistic, daily research workflows.\footnote{Our work has been reviewed by our institution's IRB.}

Participants utilized two modes of GenAI: the standard chat mode and Deep Research. We included Deep Research because it is designed for extended research tasks so it is interesting to examine whether its research-oriented capabilities benefit interdisciplinary researchers specifically. Providing both modes also allowed us to study how researchers selected and orchestrated different GenAI capabilities depending on their research needs as it reflects an ecologically valid setting in which researchers could choose the mode most appropriate for their ongoing work.

\subsubsection{Participant recruitment}
We recruited 15 participants (7 female, 7 male, 1 non-binary) through snowball \cite{goodman1961snowball, noy2008sampling} and convenience sampling, fluent in English and actively leading interdisciplinary projects. Participants provided informed consent and received a \$50 gift card.

\subsubsection{Phase 1: Pre-interaction Survey}
Participants completed a 5-minute survey covering their research background, ongoing project, AI experience, and demographics, selecting one project for the study and reporting confidence in their primary and secondary research areas.² Table 1 summarizes participants, all located in the United States, spanning computer science (26.67\%), other STEM (40\%), and social sciences (33.33\%). While 80\% reported high confidence ($\geq$8) in their primary area, only 26.67\% did in their secondary area (modal scores 3 and 5), the confidence gap motivating our focus on interdisciplinary knowledge-seeking. GenAI familiarity was generally high (60\% rated $\geq$8), but Deep Research familiarity was markedly lower (mean=4.73, mode=1).\footnote{All questionnaires are included in the Appendix.}

\begin{figure*}
    \centering
    \includegraphics[width=0.9\linewidth]{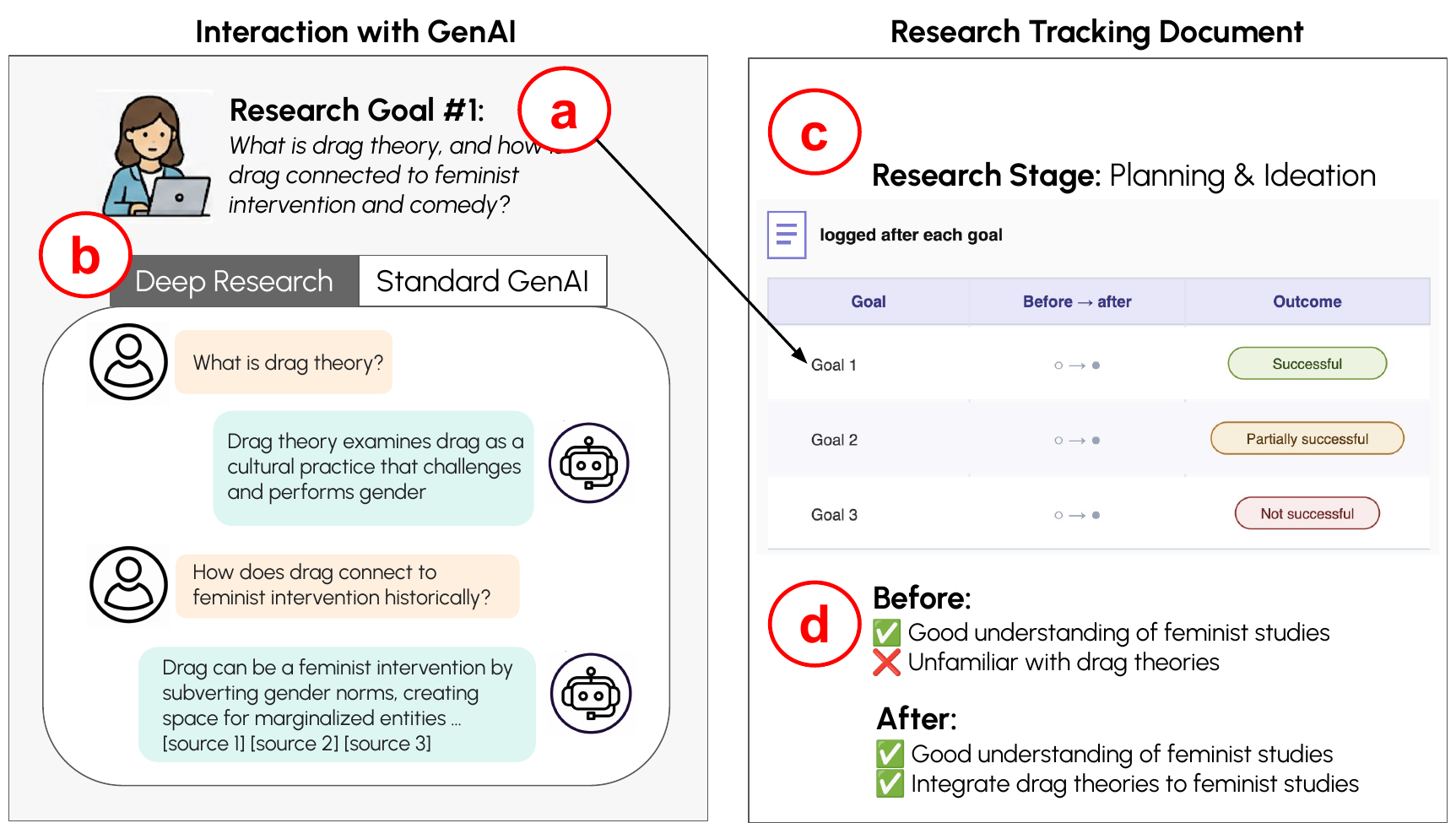}
    \caption{Interaction phase. Before interaction phase, each participant defines their research goal (a). They then interact with GenAI in either standard or Deep Research mode for 1-2 weeks (b) and log each goal and states of the work before and after interacting with GenAI (c and d).}
    \label{fig:main_figure}
\end{figure*}

\subsubsection{Phase 2: Longitudinal Interaction Period}
Participants interacted with OpenAI's standard ChatGPT and Deep Research mode over 1–2 weeks between July and September 2025.\footnote{We provided participants with credentials to access the tools.} At the time of data collection, Deep Research could ask follow-up questions before retrieving and synthesizing information into a structured, cited report. Participants were limited to five Deep Research uses per month and used standard mode freely otherwise.

A \textbf{research goal} refers to what a participant seeks to accomplish with GenAI, such as ``exploring ways to apply LLMs in mental health research and global equity.'' Participants were asked to avoid simple, generic tasks such as grammar or spelling correction that were unrelated to their research. Throughout the study, participants maintained a research tracking document (Figure \ref{fig:main_figure} c–d), recording the goal, relevant context, and the ``Before'' and ``After'' states of their work, then rating each goal as Successful, Partially Successful, or Not Successful with a brief explanation. This success rating reflects a participant's judgment of whether the interaction advanced their research goal and is not an independent measure of output accuracy. At least half of each participant's goals were required to address interdisciplinary needs.

We collected 411 prompt–response pairs (27\% involving Deep Research). Participants engaged in an average of 27.4 turns (median=19), used Deep Research an average of 2.53 times, and pursued 6.27 goals each as shown in Table \ref{tab:interaction_summary}. \footnote{Including goals inferred by authors, there are 94 distinct goals in total, but some goals were revisited later by the participants in the interaction log. The goal-level analysis (\S\ref{sec:allocate}) uses 69 goals that received a full disciplinary and task-type classification. The turn-level analysis (\S\ref{sec:orchestrate}) uses 97 goal-instances instead, since each revisit is counted as a separate instance.}

\begin{table}[]
\centering
\begin{tabular}{lccc}
\toprule
\textbf{Turn or goal} & \textbf{Mean} & \textbf{Median} & \textbf{Mode} \\
\midrule
Total turns per participant & 27.4 (17.92 without outliers) & 19 & 13 \\
Deep Research uses & 2.53 & 2 & 2 \\
Goals per participant & 6.27 & 6 & 6 \\
Goals by participant & 4.80 & 5 & 5 \\
Inferred goals & 1.47 & 1 & 0 \\
\bottomrule
\end{tabular}
\caption{Summary of participants' interactions with AI tools (N=15). We report averages with and without outliers (P2 and P4 for turns, P2 for goals).}
\label{tab:interaction_summary}
\end{table}


\subsubsection{Phase 3: Post-Interaction Interview}
After the interaction period, we conducted one-hour semi-structured interviews over Zoom.\footnote{\url{https://www.zoom.com}} Interviews focused on participants' experiences incorporating GenAI into interdisciplinary research workflows and where they perceived each mode to succeed or fall short. Participants first described their research project, then walked through one or two goals involving Deep Research and another using only standard mode. For each goal, they explained their reasoning and identified whether it concerned their primary area, secondary area, or both. We also asked participants to compare the two modes across different interdisciplinary research challenges. Interviews were recorded with consent, transcribed, and analyzed using inductive thematic analysis \cite{braun2012thematic}. Two interviewers independently conducted iterative open coding in Mural,\footnote{\url{https://www.mural.co}} with the research team meeting regularly to review emerging themes. We triangulated interview accounts with interaction logs to examine whether reported strategies were reflected in actual tool use.

\subsection{Data Analysis}
We analyzed interview transcripts, research tracking documents, and GenAI interaction logs at two levels: \textit{goal} and \textit{prompt}. At the goal level, we identified GenAI usage type and disciplinary area (primary, secondary, or both). At the prompt level, each user prompt served as the unit of analysis for characterizing how participants operationalized their research goals.

\subsubsection{Goal-Level Analysis}
We categorized participants' research goals using a taxonomy adapted from Liao ~\citep{liao2024llms}, collapsing \textit{editing} and \textit{writing} into one label and excluding \textit{data-generation} since it is outside our scope. The resulting taxonomy comprises nine categories: \textit{technical information seeking}, \textit{discover papers}, \textit{writing}, \textit{programming}, \textit{brainstorming}, \textit{summarization}, \textit{suggestion}, \textit{proof}, and \textit{data analysis}.\footnote{Definitions and examples are provided in the Appendix.}

Each goal was also coded by disciplinary relevance: \textit{primary research area}, \textit{secondary research area}, or \textit{both} (interdisciplinary). Participants wrote 69 goals. \textit{Technical information seeking} (n=26) and \textit{discover papers} (n=13) were most common, followed by \textit{programming} (n=9), \textit{brainstorming} (n=9), and \textit{writing} (n=7). The remaining categories, \textit{suggestion} (n=2), \textit{summarization} (n=1), \textit{proof} (n=1), and \textit{data analysis} (n=1), were comparatively rare. We examine how goal types relate to GenAI resource choice and outcome in \S\ref{sec:allocate_goal_type}.

\begin{figure}
    \centering
    \includegraphics[width=0.4\linewidth]{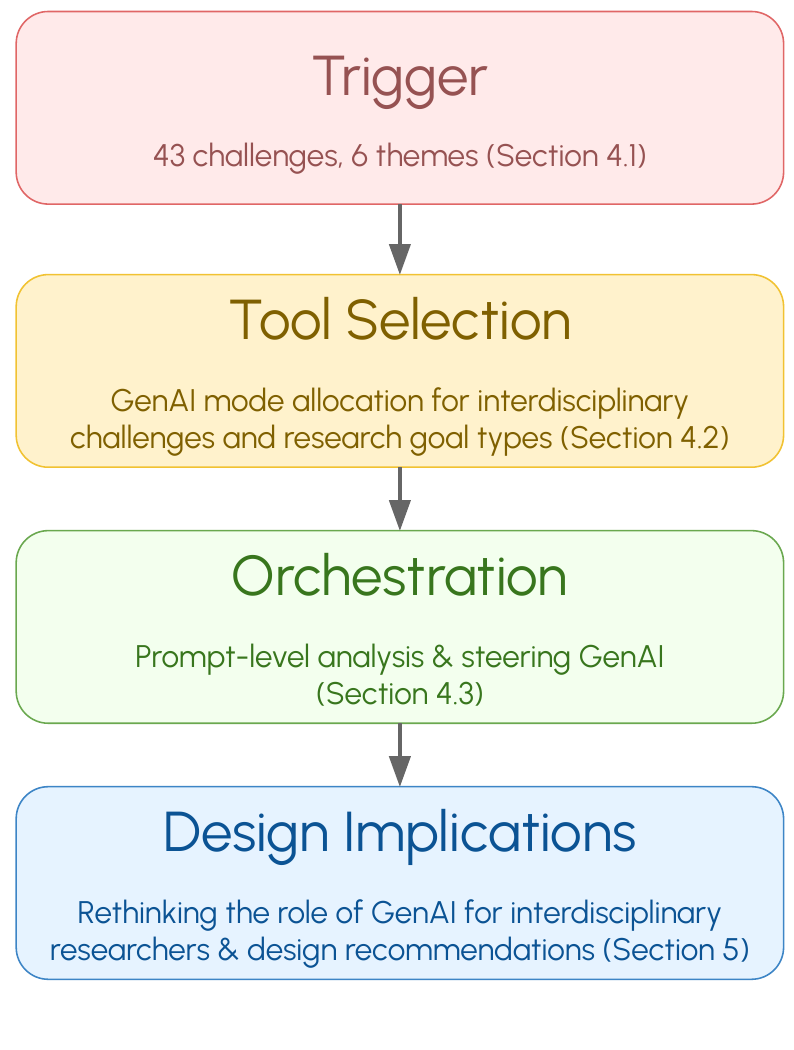}
    \caption{Overview of our findings and discussion. We first surface the interdisciplinary challenges that lead researchers to seek GenAI support (Section 4.1), then examine which GenAI mode they select for which challenges and research goal types (Section 4.2), and how they orchestrate and steer GenAI turn by turn within a conversation (Section 4.3). These inform our design implications for interdisciplinary research support (Section 5).}
    \label{fig:story}
\end{figure}

\subsubsection{Prompt-Level Analysis}
In addition to goal categories, we analyzed the conversation logs to examine how researchers operationalize their research goals through their prompts to GenAI tools. We organized the prompt-level codes into five overarching themes: \textit{Grounding}, \textit{Information-seeking}, \textit{Iterative refinement}, \textit{Knowledge Synthesis}, \textit{Miscellaneous}.\footnote{Definitions and examples of all prompt labels are provided in the Appendix.} Each prompt serves as the unit of analysis, and a prompt could receive multiple codes when it serves multiple functions; applying this scheme to our 411 tracked turns yielded 678 total code instances. Each prompt was also coded by its disciplinary relevance: \textit{primary research area}, \textit{secondary research area}, or \textit{both} (interdisciplinary).

Three of the authors initially coded the same subset of the conversation and met with the research team to iteratively develop and refine the coding scheme. After finalizing the annotation scheme, two authors independently annotated conversations of four participants, achieving 95\% inter-annotator agreement (measured by percentage agreement), indicating high reliability. The two researchers then independently coded the remaining conversations.\footnote{Details of prompt-level code statistics are provided in Table \ref{tab:prompt_stats} in the Appendix.}
\begin{itemize}
    \item \textbf{Grounding} is a process when users and systems align their knowledge states to establish sufficient mutual understanding for a shared task \cite{brennan1990seeking, clark1991grounding, brennan1996lexical, brennan2014grounding}. Codes include: \textit{declare background of research need}, \textit{provide information}, \textit{provide additional information}.
    
    \item \textbf{Information-seeking} captures user prompts that request new information to bridge knowledge gaps \cite{case2016looking, marchionini2006exploratory, pirolli2005sensemaking, choo2000information, krikelas1983information}. Codes include: \textit{ask for definition}, \textit{ask for explanation}, \textit{ask for examples}, \textit{exploration}, \textit{brainstorming}, \textit{technical question}.
    
    \item \textbf{Iterative refinement} captures how users revise prompts after receiving GenAI responses, including rephrasing, narrowing, probing, and repairing miscommunication \cite{foster2004nonlinear, schegloff1991conversation}. Codes are \textit{modify prompt}, \textit{follow-up question}, \textit{request clarification}, \textit{positive feedback}, and \textit{negative feedback}.
    
    \item \textbf{Knowledge synthesis} captures requests to integrate, validate, and structure existing knowledge, including users' own understanding, prior model responses, or known scientific literature \cite{sawyer2005cambridge, van2009distinguishing}. Codes include \textit{seeking validation}, \textit{ask LLM to analyze/process data}, \textit{ask LLM to summarize}, and \textit{ask LLM to compare between two or more methods that the user already knew}.
    \item \textbf{Miscellaneous} captures prompts that do not fit into the other categories: \textit{provide scope}, \textit{acknowledgment}, \textit{others}. 
\end{itemize}

\section{Findings}

Our findings examine the challenges that motivate researchers to seek GenAI assistance (\S\ref{sec:trigger}), how researchers allocate GenAI resources based on their needs and goals (\S\ref{sec:allocate}), and how they orchestrate GenAI through turn-by-turn interaction (\S\ref{sec:orchestrate}) as summarized in Figure \ref{fig:story}. 


\subsection{What Challenges Motivate Interdisciplinary Researchers to Seek GenAI Support?}
\label{sec:trigger}
From the fifteen post-interaction interviews, we identified 43 distinct statements related to challenges in interdisciplinary research. We organized them into six interdisciplinary needs that motivate researchers to seek GenAI support: knowledge discovery, synthesizing two research domains, methodology, framing, collaboration, and novelty. The emerging themes are shown in Table~\ref{tab:challenges}.

\begin{table*}[t]
\centering
\small
\setlength{\tabcolsep}{4pt}
\renewcommand{\arraystretch}{1.15}
\begin{tabularx}{\textwidth}{>{\raggedright\arraybackslash}X>{\raggedright\arraybackslash}X>{\raggedright\arraybackslash}X}
\toprule
\textbf{\textcolor{cKnowledge}{Knowledge Discovery \primary \secondary \integration}} (N=11) &
\textbf{\textcolor{cSynthesis}{Integration \integration}} (N=5) &
\textbf{\textcolor{cMethods}{Methodology \secondary}} (N=5) \\
\midrule
\textbf{K.I} Missing basic knowledge \textcolor{cKnowledge}{\secondary} \newline
\textbf{K.II} No shared vocabulary or taxonomy to search across the two fields \textcolor{cKnowledge}{\secondary \integration} \newline
\textbf{K.III} Inability to judge correctness or novelty \textcolor{cKnowledge}{\secondary \integration} \newline
\textbf{K.IV} Fragmented literature/data infrastructure \textcolor{cKnowledge}{\primary \secondary \integration}
&
\textbf{I.I} Connecting two areas not normally bridged \textcolor{cSynthesis}{\integration}\newline
\textbf{I.II} Generating a new angle after getting stuck on the same material \textcolor{cSynthesis}{\integration}
&
\textbf{M.I} Selecting or validating the right method/tool version \textcolor{cMethods}{\secondary} \newline
\textbf{M.II} Picking up new technical skills/methods outside training \textcolor{cMethods}{\secondary}
\\
\multicolumn{3}{c}{} \\[-8pt]
\toprule
\textbf{\textcolor{cFraming}{Framing \integration}} (N=2) &
\textbf{\textcolor{cIsolation}{Collaboration \secondary \integration}} (N=5) &
\textbf{\textcolor{cNovelty}{Novelty \primary}} (N=3) \\
\midrule
\textbf{F.I} Same term means different things in each field \textcolor{cFraming}{\integration} \newline
\textbf{F.II} Discovering a field's perspective to frame the problem \textcolor{cFraming}{\integration} \newline
\textbf{F.III} Repeated rejection tied to unmastered cross-disciplinary framing \textcolor{cFraming}{\integration} 
&
\textbf{C.I} No collaborators with the needed expertise \textcolor{cIsolation}{\integration} \newline
\textbf{C.II} Skills gap surfaced only through interpersonal friction, not self-detection \textcolor{cIsolation}{\secondary}
&
\textbf{N.I} Original theorems/proofs too novel to delegate \textcolor{cNovelty}{\primary} \newline
\textbf{N.II} Synthesizing artifacts with no literature precedent at all \textcolor{cNovelty}{\primary}
\\
\bottomrule
\end{tabularx}
\caption{Six themes of interdisciplinary researchers' challenges identified in this study, with the number of participants (N) who expressed each struggle during the post-interaction interview. \primary indicates that the struggle arose in the participant's primary research area, \secondary = secondary research area, and \integration = when the participant tries to integrate both areas.}
\label{tab:challenges}
\end{table*}
\subsubsection{Challenge 1: Knowledge Discovery}
Knowledge discovery challenges happen as researchers navigate an unfamiliar domain. Most participants (N=11) described challenges related to knowledge discovery most often in the secondary area \cite{daniel2022challenges}. 
Participants described building baseline knowledge in secondary fields (P3, P6, P7, P14), identifying unfamiliar search vocabulary (P1), and judging information they were less equipped to verify (P5, P15). They also struggled to identify the vocabulary needed to search across fields, as when P1 was unsure which keywords would surface literature at the intersection of mixed reality and robotics. Others could not judge what they found: P5, new to virology, described being ``\textit{not a good verifier of the area}.'' Finally, the infrastructure itself was sometimes the obstacle, as P10 noted that no central database tracks pathway-disease associations in human genetics. P2 captured the shared experience:
\begin{quote}
\textit{``Sometimes I have very basic questions in my mind because of the interdisciplinary nature of my research. I have good knowledge in my field, but I might not have all the knowledge in other fields. I have to go back to search for basic information...''} — P2
\end{quote}

These challenges arose primarily when participants pursued goals in their secondary fields. The difficulty is not the availability of information but the need to acquire it quickly, raising the question of how GenAI can help interdisciplinary researchers become sufficiently informed to navigate a secondary field.

\subsubsection{Challenge 2: Integration}
Interdisciplinary challenges related to integration challenges involve combining insights from multiple research areas rather than simply acquiring knowledge from a secondary field. Prior HCI research has similarly characterized interdisciplinary collaboration as requiring researchers to negotiate and connect different disciplinary perspectives, concepts, and practices \cite{wan2026knit, lee2026embracing}. For P1, P9, and P11, this meant connecting concepts that are rarely bridged: P11 sought to link cultural sociology concepts with computational methods that do not naturally map onto them, and P9, who studies transnational feminism and humor studies, needed to relate drag theory to comedy:
\begin{quote}
\textit{``One question [that I have] is that how we can relate comedy with drag theory...  But I don't have enough information or ideas about drag theory and also in relation to comedy... drag in feminist and queer theory is really important ''} — P9
\end{quote}
P9 also used GenAI to obtain a new angle after becoming stuck with the same material. Unlike knowledge discovery, integration required researchers to determine how concepts or approaches from one domain could inform another.

\subsubsection{Challenge 3: Method}
Five participants faced method or skill gaps in the secondary domain they were less trained in; here they had a clear goal but lacked the methodological knowledge to carry it out. Some needed help with execution as P2 had to repeat an analysis after realizing a package version had affected her results. P4 found it useful that GenAI laid out alternatives and when to prefer each. Others turned to GenAI to learn a secondary field's practices. For instance, P8 used GenAI for data analysis because she was not confident with the tools needed to interpret her results while P12 said: ``\textit{I needed to learn from labor literature how they are doing their measurements at the aggregated level… from reports like Bureau of Labor Statistics.}'' This use of GenAI as a means of learning a field's practices, rather than simply executing a task within it, echoes prior work supporting novice researchers in building unfamiliar research skills with language models \cite{choe2024supporting}.

\subsubsection{Challenge 4: Framing}
Framing challenges arise when researchers must determine how a concept or research question should be understood from another field's perspective, often because the same term carries different meanings across fields. P3 described how GenAI prompted her to reconsider her research from another field's standpoint: \textit{``It also makes me realize that we need to think from other perspectives… I wouldn't have the motivation to think from the other side.''} For P12, framing was tied to communicating research in a way that met the secondary field's expectations; after repeated paper rejections they identified a gap in their understanding of labor economics that had shaped how they framed the work. Interdisciplinary framing thus involves not only deciding \textit{what} to study, but which disciplinary perspective and terminology make the research legible at the intersection.

\subsubsection{Challenge 5: Collaboration}
P4, P11, and P15 described difficulty accessing collaborators with the expertise their interdisciplinary work required: P4's collaborators lacked sufficient data science expertise, P11 had few nearby scholars working on Chinese-context research, and P15 depended on lab members with relevant experience when troubleshooting experiments. P2 described a related challenge in which a gap in her own knowledge surfaced only through interaction with a collaborator: \textit{``He was super angry with me last time when I met him. I didn't know the difference between the packages.''} Interdisciplinary researchers therefore face challenges not only in acquiring expertise, but in identifying and accessing the expertise needed to move their research forward. In these situations, participants could use GenAI as an accessible source of guidance when relevant human expertise was unavailable or difficult to access.

\subsubsection{Challenge 6: Novelty}
P6, P7, and P8 described challenges arising when their work had little or no existing precedent for GenAI to draw on. P6 and P8, both developing and proving new theorems, found the complexity and specificity of their formulations difficult to delegate, and P8 noted that the many concepts introduced in his own work made new theorem statements hard to simplify. As P6 explained, \textit{``If we want to correctly prove the conclusion, we might have to discuss in like 8 pieces... One of them is there's a lot of small mistakes that will break everything.''} Unlike the other struggles, novelty challenges arose in participants' primary research areas. The difficulty was not acquiring knowledge from an unfamiliar field, but working on problems for which existing literature offered limited guidance. 

\subsection{How Do Researchers Allocate GenAI Assistance Across Research Goals?}
\label{sec:allocate}
To understand how interdisciplinary researchers allocate GenAI assistance, we examine two dimensions: the research need identified in \S\ref{sec:trigger} and the goal type \cite{liao2024llms}, which captures the nature of the assistance sought (e.g., information seeking, brainstorming). We then relate these goal characteristics to researchers' selection of GenAI modes (Deep Research vs. standard GenAI) and to the outcomes of their interactions.


\begin{figure}
    \centering
    \begin{subfigure}[b]{0.95\textwidth}
        \centering
        \includegraphics[width=\textwidth, trim=0 1cm 0 7cm,clip]{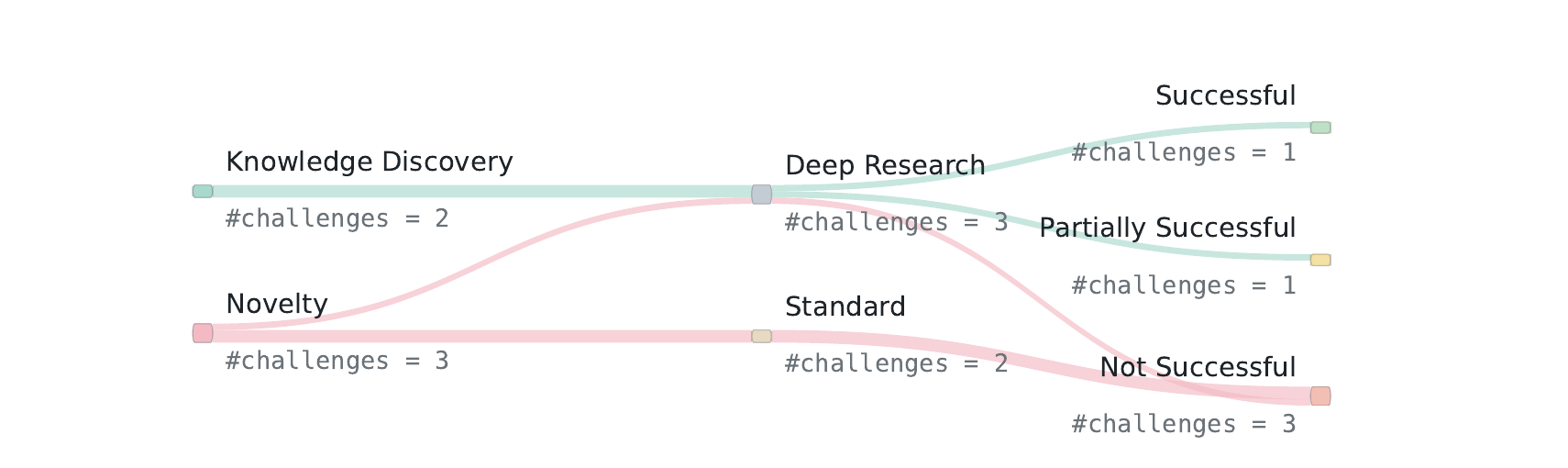}
        \caption{Primary discipline}
        \label{fig:workflow-primary}
    \end{subfigure}
    \begin{subfigure}[b]{0.95\textwidth}
        \centering
        \includegraphics[width=\textwidth, trim=0 1cm 0 7cm,clip]{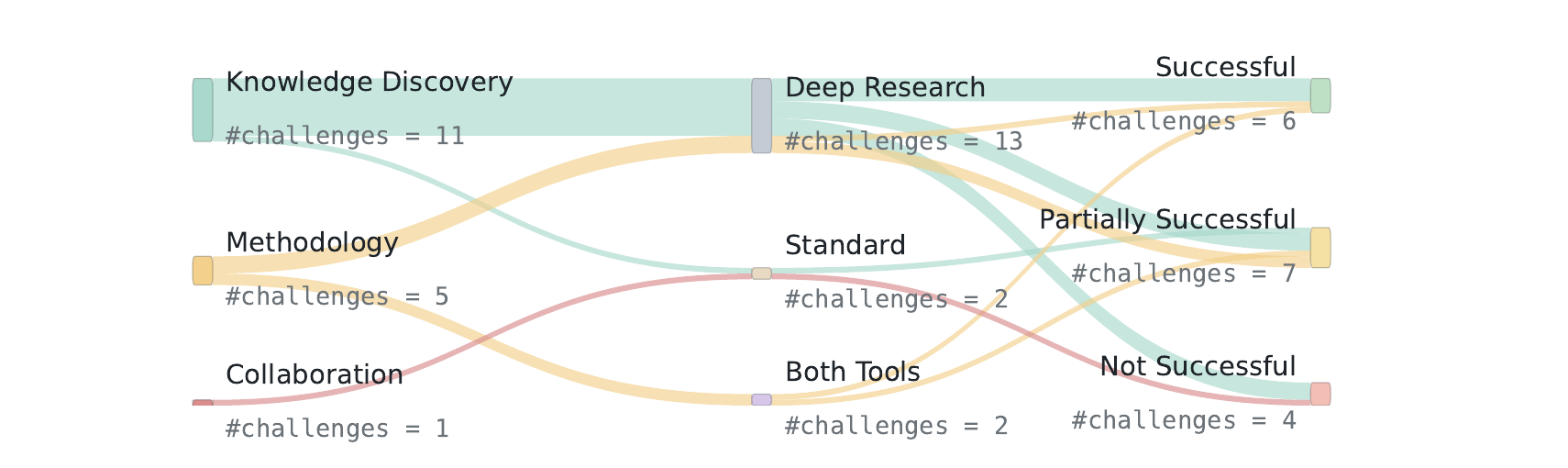}
        \caption{Secondary discipline}
        \label{fig:workflow-secondary}
    \end{subfigure}
    \begin{subfigure}[b]{0.95\textwidth}
        \centering
        \includegraphics[width=\textwidth, trim=0 1cm 0 6.7cm,clip]{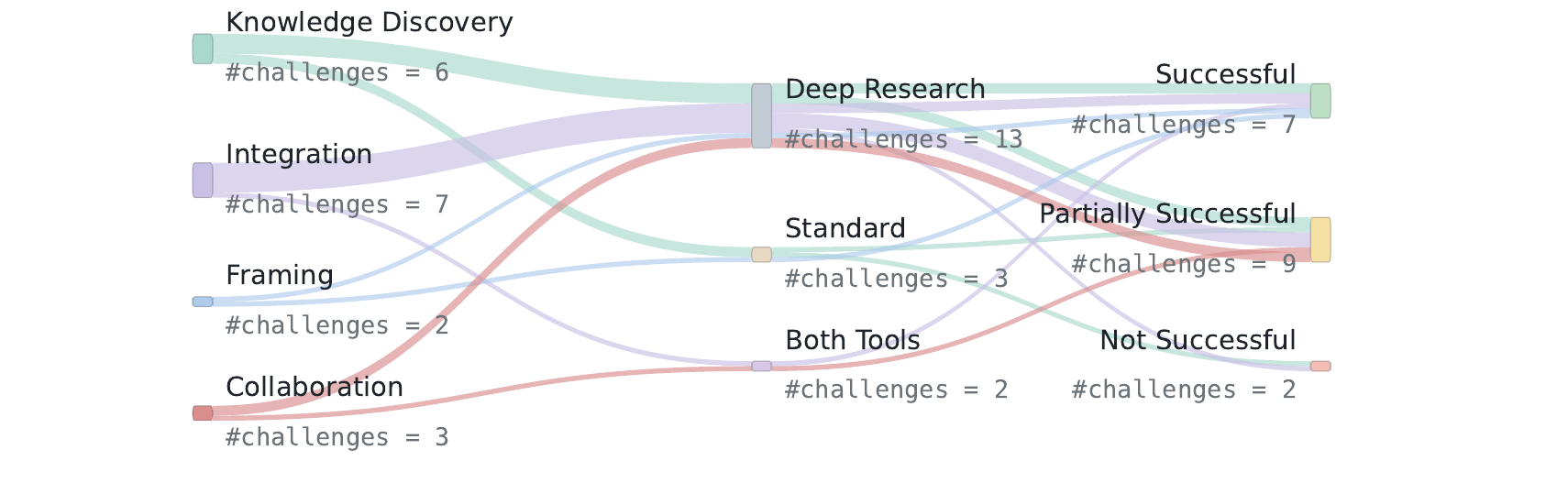}
        \caption{Both disciplines}
        \label{fig:both-challenges}
    \end{subfigure}
    \caption{Three sankey diagrams showing which GenAI (Deep Research, standard, or both tools) was used for each goal type and whether GenAI's assistance was rated successful, partially successful, or not successful.}
    \label{fig:genai_for_research_need}
\end{figure}

\subsubsection{GenAI Allocation for Interdisciplinary Research Challenges}
\label{sec:allocation_for_research_challenges}
To examine GenAI selection to address the interdisciplinary research challenges, we triangulated (1) interviews, (2) research tracking documents, and (3) chat logs with GenAI. Figure \ref{fig:genai_for_research_need} presents a Sankey diagram tracing research goals associated with 40 of the 43 challenge statements identified from participants' interviews across three stages: interdisciplinary challenge, GenAI tool mode (standard GenAI or Deep Research), and outcome.\footnote{Three challenges mentioned by participants did not lead to GenAI use so they are not included in this analysis.}

\paragraph{Deep Research as an entry point.} Deep Research was the default destination for challenges related to knowledge discovery and methodology, as most of such goals were routed to Deep Research. It suggests that participants expects broader search could compensate for knowledge gaps. Participants described Deep Research as a good starting point (P1, P3, P4, P5, P8, P14) as it could surface keywords and papers. They also appreciated that Deep Research could not only output relevant papers but also organize results into thematic summaries (P2, P5, P10, P12, P14). As P2 noted, ``it helps me stay focused to my problems, rather than lost in the sea of manuals.'' Deep Research also produced \textbf{fast justifications for methodological choices and designs (N=6)}, and sometimes hypotheses that researchers carried into their own writing (P1, P6, P7, P11, P13, P15). P15, for instance, obtained justification from her secondary field (cellular biology) for a result in her primary field: ``\textit{it took the primary literature and hypothesized for me about why I might be seeing that effect... I actually used to contribute to my manuscript...  I use this to influence my discussion.}'' Similarly, P11 benefited from diverse explanations of computer science approaches to process data within their primary field of cultural sociology, while P13 leveraged AI to explore their goal of determining how AI might be used to address the ongoing problem of data decay.


\paragraph{Integration requires the researcher's framing.} 
Seven integration goals were routed to Deep Research, yet participants noted that \textbf{the tool struggled to connect across domains unless they explicitly framed the prompt to do so}. For instance, P7 could connect chemistry and biology only through prompts that emphasized how molecules interact differently within cells and animals; P9 similarly found that only detailed questions led the tool to integrate drag theory and comedy. In all these cases, researchers are the ones who make interdisciplinary connections by explicitly framing their
prompts to include both domains, rather than relying on Deep Research to bridge them on its own. When Deep
Research attempted to make connections, P11 noted that these links felt forced and unintuitive, reflecting more the
tool’s interpretive stretch than genuine disciplinary overlap. In such cases, Deep Research was seen as unable to produce
coherent interdisciplinary outline.

\paragraph{Novelty cannot be retrieved.} All three novelty-related goals were rated Not Successful regardless of GenAI mode. The barrier is not retrieval breadth but the absence of prior work to retrieve. P7, a synthetic chemist, explained that ``\textit{I'm making the molecules that nobody else has made before, so I think in this case, chatGPT cannot provide super useful suggestion,  and that is the majority of my work.}'' Deep Research's more extensive search offered no advantage over standard chat because existing literature has nothing to surface.

\begin{figure}
  \centering
  \begin{subfigure}[t]{0.49\textwidth}
    \centering
    \includegraphics[width=\textwidth]{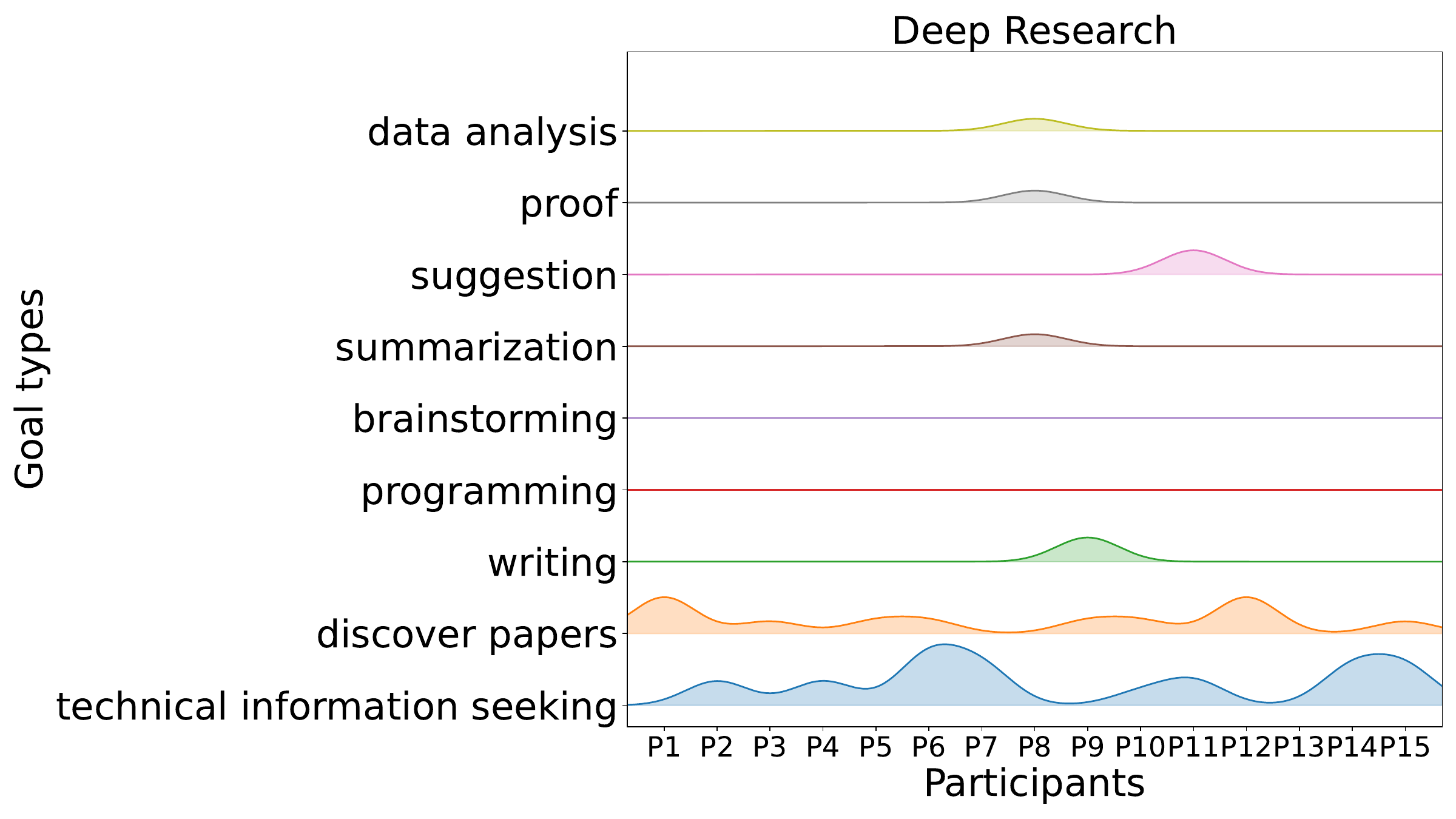}
    \caption{Deep Research goal types}
    \label{fig:subfig1}
  \end{subfigure}
  \hfill
  \begin{subfigure}[t]{0.49\textwidth}
    \centering
    \includegraphics[width=\textwidth]{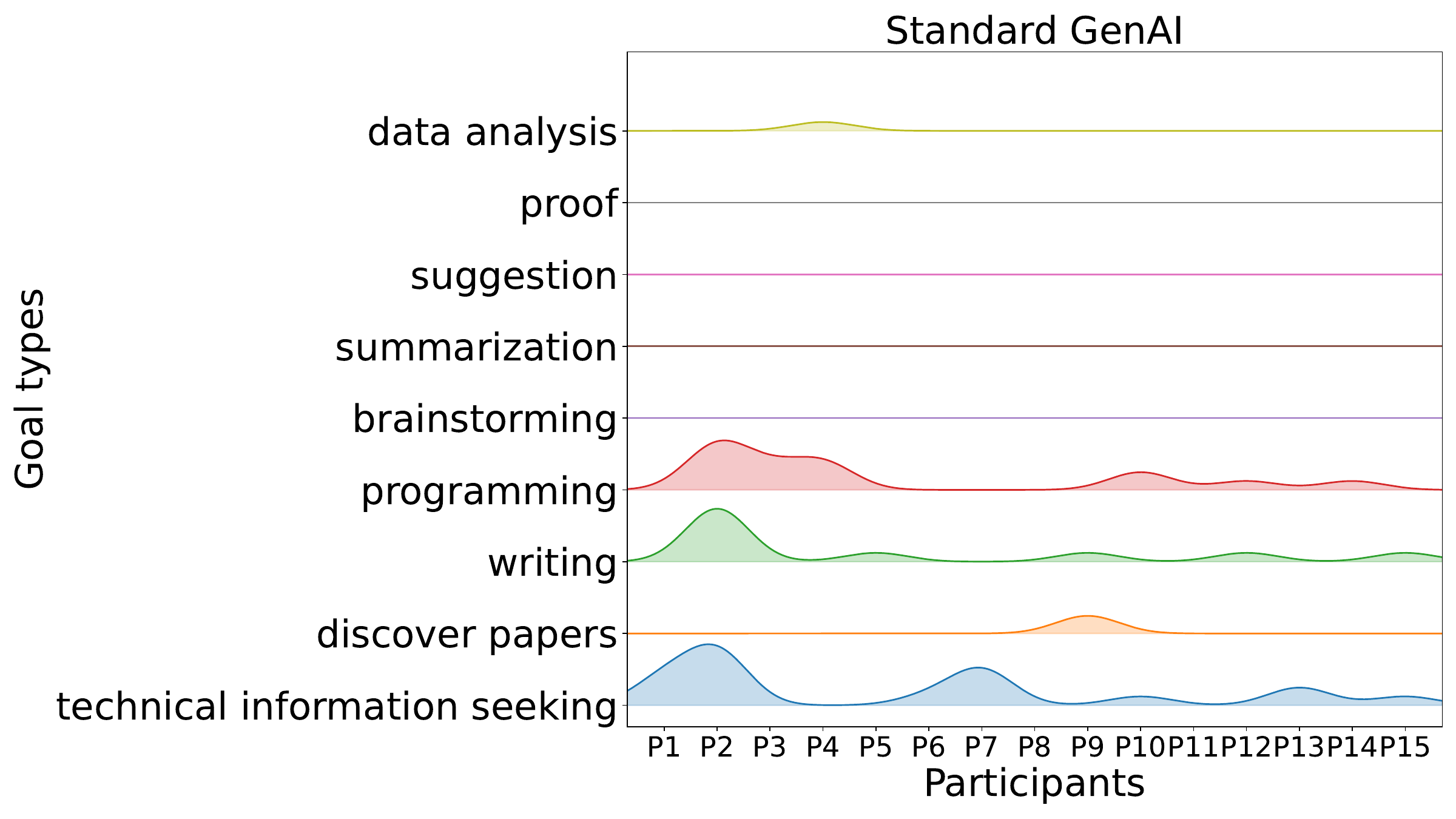}
    \caption{Standard GenAI goal types}
  \end{subfigure}
  \caption{Distribution of goal types across participants (P1–P15) in (a) Deep Research and (b) standard GenAI sessions.}
  \label{fig:goals_deep_research_vs_genai}
\end{figure}
\subsubsection{GenAI Allocation Depends on the Research Goal}
\label{sec:allocate_goal_type}
We next examined how researchers allocated GenAI across 69 tracked goals.\footnote{The 69 goals include all participant-tracked goals, whereas the analysis in §4.2.1 includes only the 40 goals associated with interdisciplinary challenges mentioned by participants in the interview and for which they used GenAI.}  As shown in Figure~\ref{fig:goals_deep_research_vs_genai}, tool choice varied by goal type. Deep Research was used more often for technical information seeking and literature discovery, whereas programming and writing were handled primarily with standard GenAI. Idea generation also leaned toward standard GenAI, where researchers could develop and refine ideas through iterative interaction.

These choices reflected what researchers expected from each GenAI mode. Deep Research was particularly useful when researchers wanted to gather information broadly or quickly build background knowledge on an unfamiliar topic. As P3 explained, Deep Research ``\textit{would gather articles online and then output a long document. So it's not like [it tries] to come up with new things -- its main focus is to synthesize things}'', and therefore defaulting to standard GenAI for brainstorming.  Deep Research is found useful for surfacing relevant information as a long report, although sometimes the output was not directly applicable for the research.

Standard GenAI, in contrast, was preferred for goals that benefited from iterative interaction. Researchers used it to debug code, revise text, brainstorm ideas, and progressively refine responses through follow-up prompts. This allowed researchers to adjust direction as their needs changed, consistent with prior work characterizing conversational GenAI as a complement to specialized search tools ~\cite{yen2024search}. 

Outcomes generally aligned with these allocation patterns although several goal-tool combinations had small sample sizes. For technical information seeking, 41\% of 17 Deep Research goals were rated successful, compared with 3 out of 9 uses of standard GenAI. For idea generation, 2 standard GenAI goals were rated successful, compared with only 1 out of 5 Deep Research goals. This aligns with prior HCI work showing that ideation and co-writing can benefit from back-and-forth interaction~\cite{dhillon2024shaping}. Given the small number of goals in several categories, we treat these outcome differences as descriptive rather than comparative evidence of tool effectiveness.

More broadly, the allocation patterns and participants' explanations suggest affordance matching as researchers developed expectations about what breadth-oriented and conversational assistance were each useful for and allocated GenAI resources accordingly. Effective orchestration therefore involved matching the demands of a research goal with the perceived affordances of different GenAI resources, rather than selecting a single ``best'' tool. This suggests that \textbf{research-oriented GenAI systems could make their affordances and limitations more visible}, helping researchers make these choices deliberately rather than through trial and error~\cite{kittur2013costs}.

\subsection{Orchestrating GenAI for Interdisciplinary Research}
\label{sec:orchestrate}
\subsubsection{How does the conversation unfold turn by turn?}
\begin{figure}[t]
    \centering
    \begin{subfigure}[b]{0.45\textwidth}
        \centering
        \includegraphics[width=\textwidth, trim=0.1cm 4cm 0.2cm 3cm,clip]{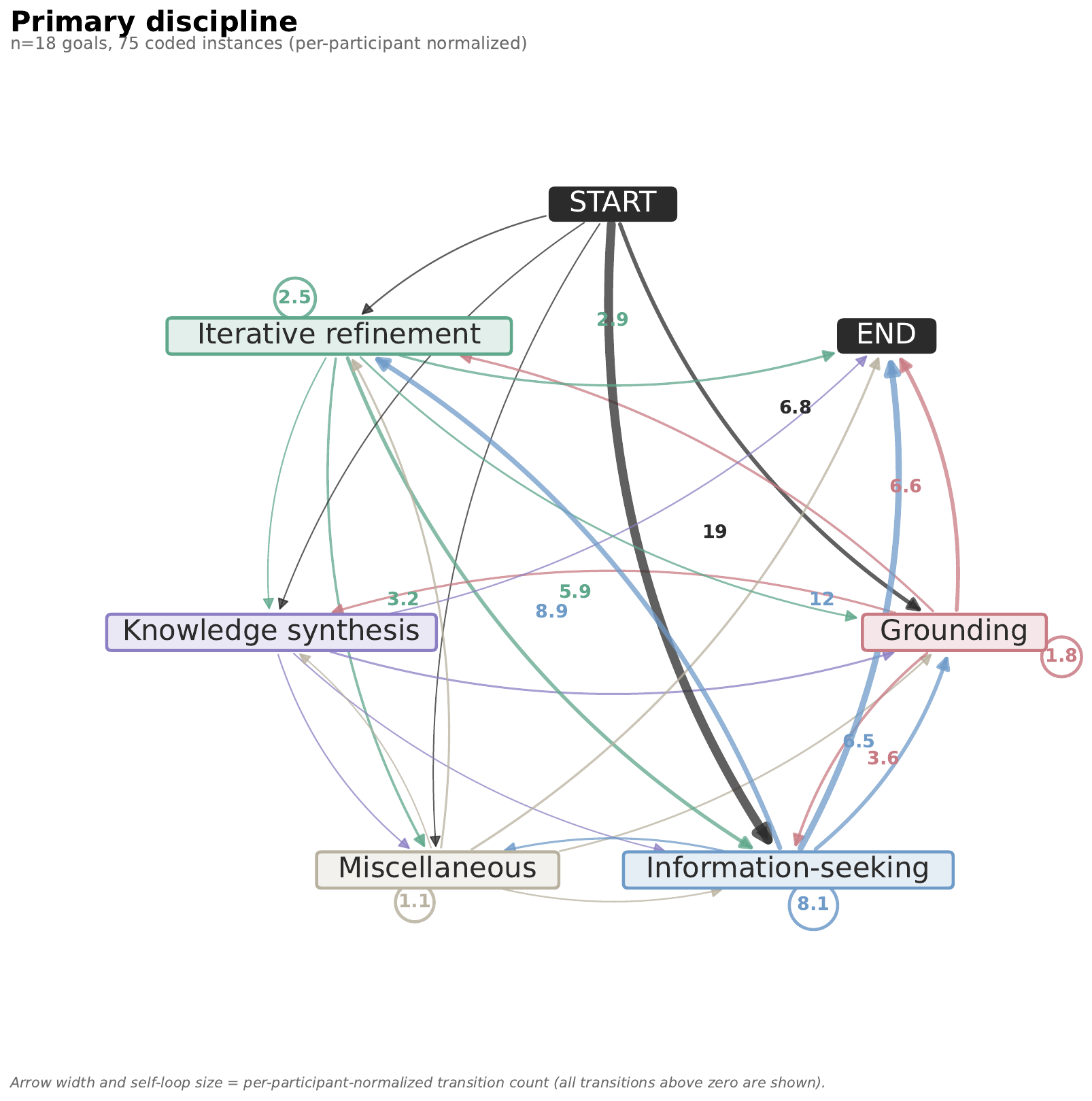}
        \caption{Primary discipline (18 goals, 75 coded instances)}
        \label{fig:workflow-primary}
    \end{subfigure}
    \hfill
    \begin{subfigure}[b]{0.45\textwidth}
        \centering
        \includegraphics[width=\textwidth, trim=0.1cm 4cm 0.2cm 3cm,clip]{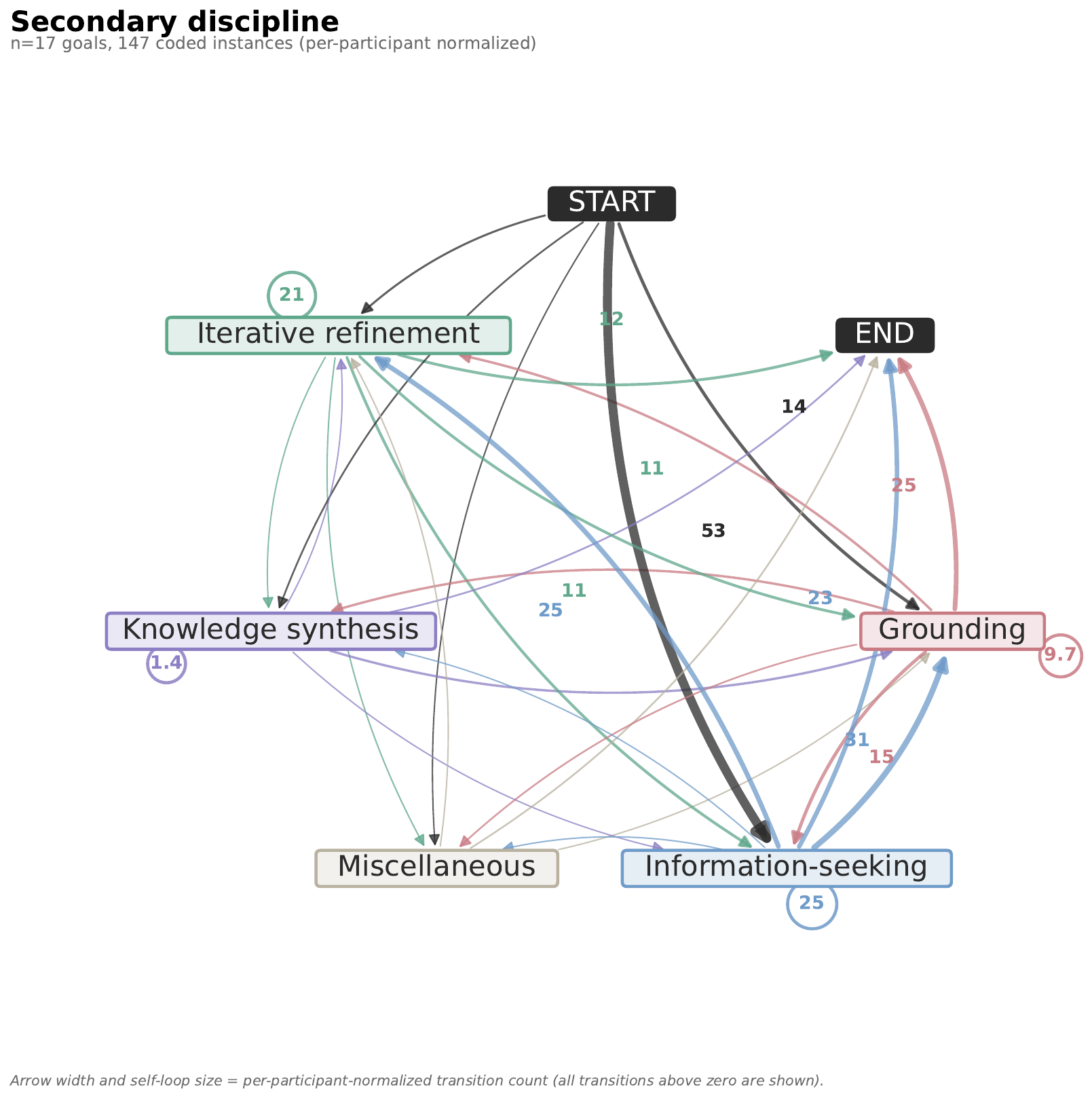}
        \caption{Secondary discipline (17 goals, 147 coded instances)}
        \label{fig:workflow-secondary}
    \end{subfigure}
    \hfill
    \begin{subfigure}[b]{0.7\textwidth}
        \centering
        \includegraphics[width=\textwidth, trim=0.1cm 4cm 0.2cm 3cm,clip]{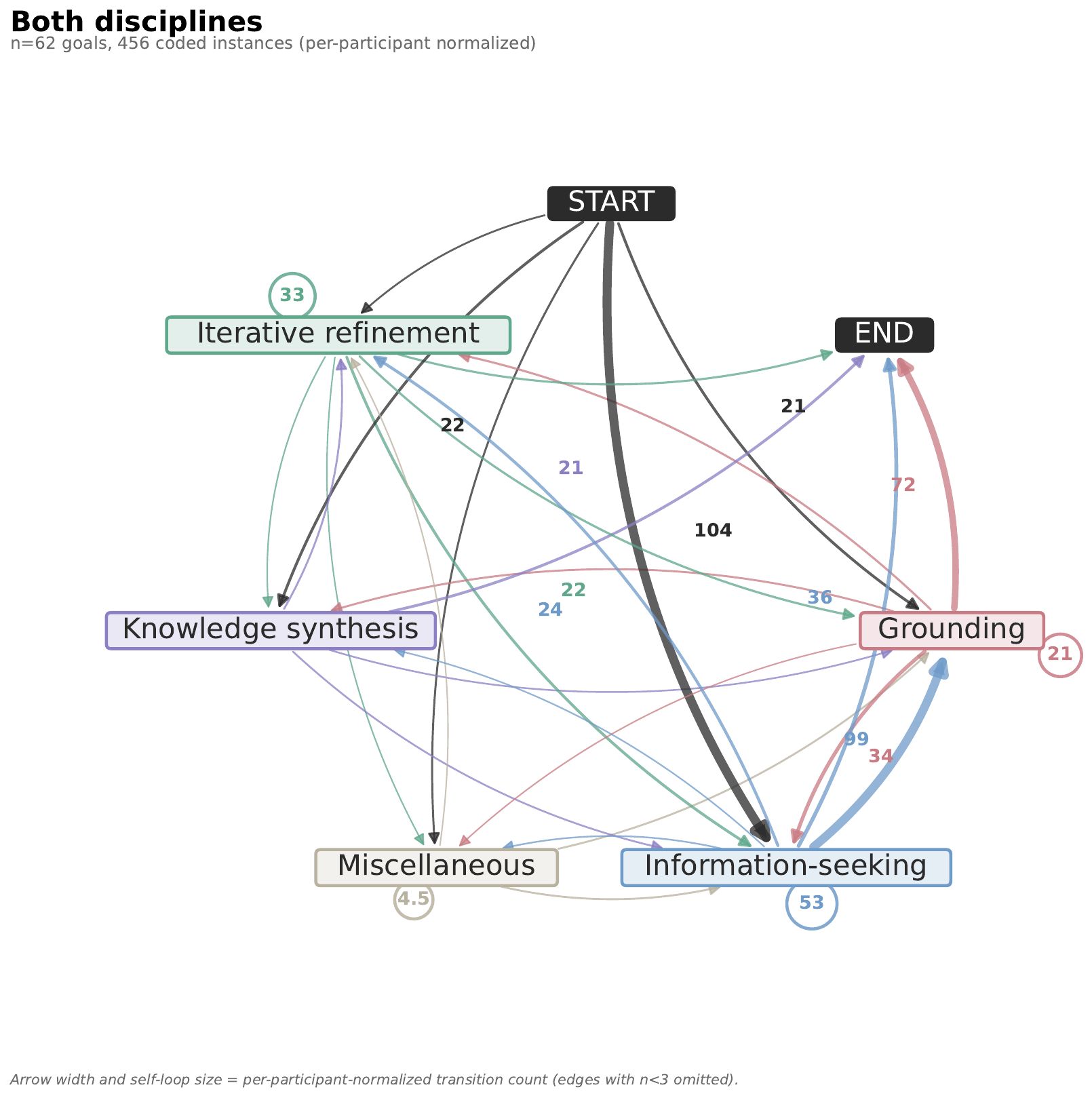}
        \caption{Both disciplines (62 goals, 456 coded instances)}
        \label{fig:workflow-both}
    \end{subfigure}
    \caption{Prompt-level code transitions by disciplinary context. Each arrow represents a transition between two consecutive prompt-level codes within a goal. Thicker arrows indicate more frequent transitions. Numeric labels show the 10 most frequent transitions. Counts are per-participant-normalized to prevent outliers from dominating the pattern.}
    \label{fig:workflow_diagram_new}
\end{figure}

The goal-level analysis shows researchers' choice of GenAI mode, but not how they used it during the interaction. Thus, we analyzed the sequential transitions of prompt-level categories across 97 goals (Figure~\ref{fig:workflow_diagram_new}).\footnote{The 97 goals include the 69 researcher-tracked goals used in the goal-level analysis as well as additional goals identified by the authors during prompt-level coding. Because this analysis focuses on interactions at the turn level, we include both types of goals.} We normalize the transition counts before aggregating since P2 and P4 logged substantially more turns (details in Appendix~\ref{app:normalization}). The three transition graphs show the transition of one prompt to the next prompt for the first research area (Figure \ref{fig:workflow-primary}), secondary area (Figure \ref{fig:workflow-secondary}), and both areas (Figure \ref{fig:workflow-both}). 

Regardless the disciplines, participants often started the conversation with information-seeking and grounding. For conversations involving the secondary disciplines and both disciplines, the most frequent cross-category transition was information-seeking to grounding. After obtaining new information, researchers returned to establishing context with GenAI to build mutual understanding. Grounding thus occurred not only at the goal start, but also as a way to contextualize new information and steer subsequent interaction \cite{cho2020role, li2025beyond}. 

The dominant self-loops also differed by disciplinary context. For secondary-area and interdisciplinary prompts information seeking and iterative refinement were the most prominent. This suggests that researchers repeatedly refined GenAI responses when working in a less familiar domain. In contrast, primary-area prompts showed repeated information-seeking but relatively little iterative refinement. 

From this pattern, we can see that researchers engage in \textbf{progressive sensemaking}, but differently depending on their disciplinary context. They iteratively corrected interpretations, requested alternative explanations, or extended emerging ideas, particularly when navigating their secondary area. These sequential patterns reveal how researchers developed and contextualized information over multiple turns, which is not visible from goal-level tool selection alone. It is consistent with prior work describing AI-supported knowledge work as an iterative process in which people reshape AI outputs in relation to their goals and knowledge \cite{barke2023grounded,liu2024selenite,yun2025generative}.

\subsubsection{Steering GenAI toward the Researcher's intent}
Participants noted that it requires active steering to obtain useful responses out of Deep Research and GenAI. 



\paragraph{Supported steering.} Steering succeeded when participants were specific and deliberate, and when they could bring their own knowledge. P9 attributed success to prompt quality: \textit{``It depends on how you ask the questions or give very clear and detailed prompts to Deep Research, then you can get very relevant answers... the control, it depends on you.''} P6 and P12 similarly described follow-up questions and narrower search terms as important while P15 found that reframing a narrowly factual question about a protein interaction as a mechanistic \textit{``why''} question produced a better answer to shape her manuscript's discussion. Compared to the turn-level analysis, this is exactly the behavior underlying three frequent self-loops. Successful steering happens not from one well-crafted prompt, but several attempts. Participants also filtered outputs using their own expertise for repurposing scope results for related work (P3), judging which suggested methods were worth pursuing (P4), separating usable leads from implausible ones (P13), or catching a Deep Research summary that attributed to a real paper a finding it did not contain (P8). Filtering was treated as part of the researcher's job, not a failure of the tool.

\paragraph{Limited steering.} The recurring limit was that control was coarse. P1 captured this: the system was \textit{``too automatic... I have a little bit [of] control, but not a very strong sense of control. Because the answer would directly give you a bunch of information, and you cannot control it at a very detailed level.''} For instance, P15 was offered only three bulleted directions, and P11 described being able to control the direction of an interaction but not the answer. Even GenAI was sometimes resistant to correction. For example, P10 asked it to limit results to human studies and received work on mice, and P13 felt her control ``\textit{loosen}'' precisely when she asked for specific numbers. P7 described a subtler GenAI resistance of an unwillingness to admit its limits: \textit{``you will never say I don't know, I'm sorry, but you keep giving you some, maybe not very useful information.''} Participants worried about the confidence with which unsupported claims were delivered, a risk for researchers in the secondary domain they are less familiar with. 

\paragraph{When can breadth be useful?} Researchers noted that the breadth of responses from Deep Research could be useful if \textbf{it is providing background knowledge and context.} P1, for example, appreciated how the tool synthesized cross-disciplinary histories of mixed reality and robotics: \textit{``it can help you understand... some history background and how they develop each other and how they are relevant... and go together to form a new discipline. So background information may help you understand the reason behind.''} Others highlighted its ability to situate specific topics within broader intellectual traditions, such as connecting drag to feminism (P9) or breadth sometimes sparked new ideas or offered useful comparative perspectives. P9, for instance, described drawing on research from adjacent cultural contexts even when it was not a direct match for her project on Iranian women's comedy: \textit{``there is some research in Pakistan and India which are related to Iran... they are the similar cultural context, but again, I can get a general idea from them.''}

\section{Discussion}

\subsection{The Expertise Paradox}
GenAI was most helpful when participants knew the least. While 80\% reported high confidence ($\geq$8/10) in their primary area, only 26.67\% did so in their secondary area, where knowledge discovery was the most common reason for seeking GenAI support.

Participants described GenAI's value in terms of this knowledge gap. In her primary area of operations management, P12 found `\textit{not much new information},'' whereas in labor economics, her secondary area, the information from GenAI was `\textit{also educating me}.'' Similarly, Deep Research surfaced little new information for P11 in her established field, but helped her enter NLP by identifying unfamiliar concepts: ``\textit{after I know there's a thing, I can do some search.}''

This complicates a common expectation in human-AI collaboration that expertise enables users to benefit from AI because they can recognize and correct imperfect outputs \cite{chen2024learning}. Our participants instead illustrate a paradox of expertise. The potential value of assistance may be greatest where users have less knowledge \cite{chen2025ai}, while that same knowledge gap makes evaluating AI-generated information harder \cite{hou2021expert}. Moreover, our participants were not complete novices, but expert researchers entering a secondary field. Their adjacent research expertise may provide some foundation for evaluating unfamiliar information \cite{gajos2022people}.

Thus, it does not suggest that GenAI substitutes for domain expertise. While GenAI provided breadth, such as entry points, keywords, and plausible directions, researchers still have to retain judgment over what to accept, adapt, or discard. The expertise paradox describes where GenAI is most \textit{useful} but since it fails to be the most \textit{trustworthy} where its value was greatest in secondary fields, precisely where researchers had less ability to verify its outputs. GenAI as research assistants ideally must provide stronger verification support when researchers work outside their primary expertise.

\begin{figure}
    \centering
    \includegraphics[width=0.85\linewidth]{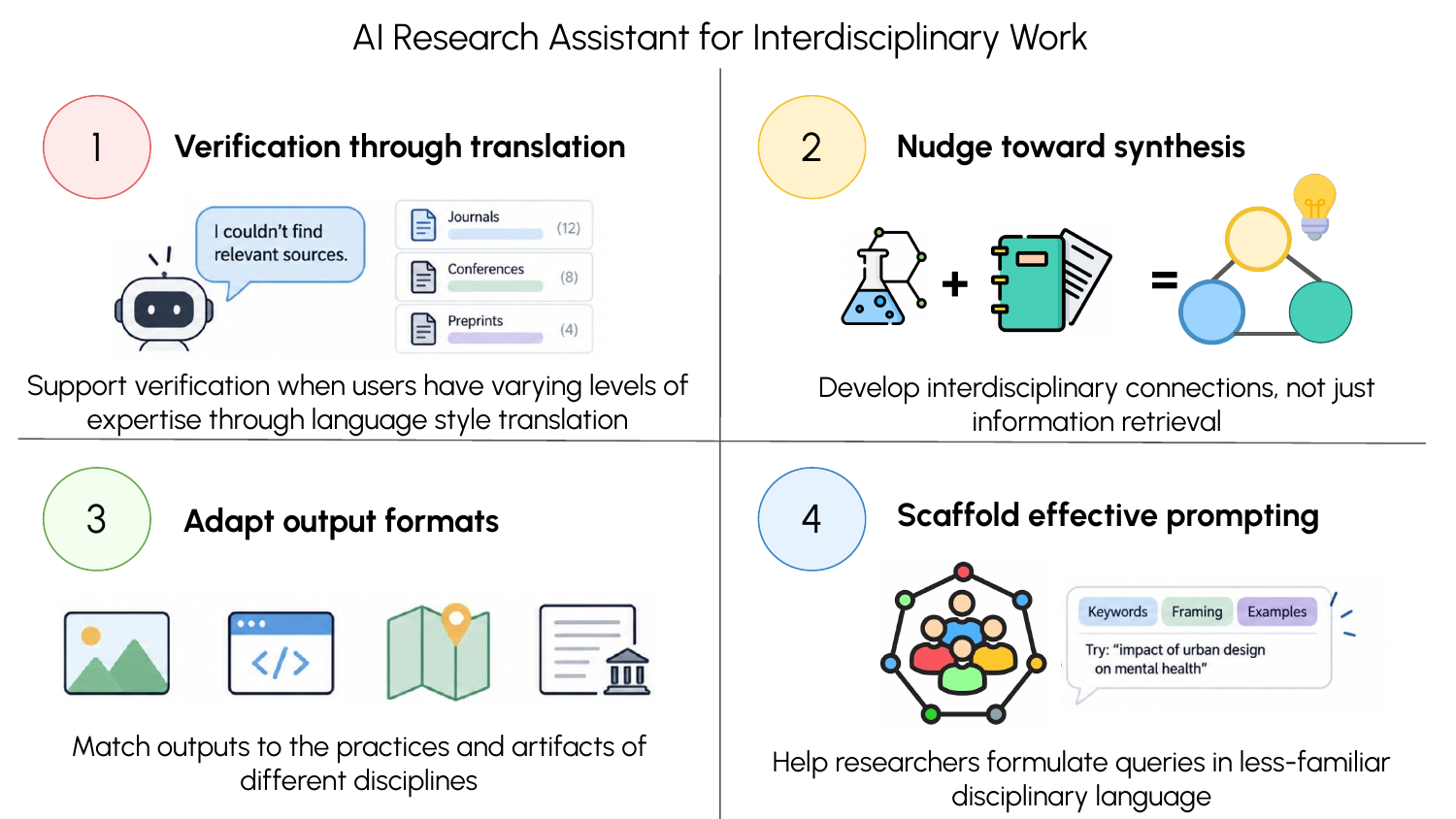}
    \caption{Design implications}
    \label{fig:design_implications}
\end{figure}
\subsection{Implications for Design}
We also asked participants what features or improvements they expected from AI research assistants for interdisciplinary work as summarized in Figure \ref{fig:design_implications}.

\paragraph{Recommendation 1: Support verification by translating into language researchers already trust.}
When researchers work in secondary domains, verification support becomes critical because they often lack the expertise to quickly judge whether a claim is well-supported. Current systems, however, may produce seemingly high-quality responses even when they cannot find relevant materials. P14 said: ``\textit{if Deep Research can't find anything relevant, then directly telling me would be better than returning a bunch of irrelevant things}''. P5 extended this perspective by mentioning: ``\textit{Don't exaggerate facts.}'' As studied by Kim et al.~\cite{kim2024imnotsure}, expressing uncertainty rather than confidently asserting an answer reduces overreliance and miscalibrated trust in LLM-based tools.

Yet communicating uncertainty alone assumes researchers can act on that signal. This assumption is weaker precisely when researchers are working outside their expertise. Our findings suggest a complementary direction: researchers are more confident and critical when evaluating work in their \textit{primary} domain since they already know what to interrogate (e.g., sample size, controls, replication, scope of claims). \textbf{GenAI tools could leverage this asymmetry by identifying the underlying method or evidentiary structure of a secondary-domain finding (e.g., a small-N qualitative study, an observational correlation, a single-cohort trial) and relating it to familiar methods in the researcher's primary field}. This could help researchers identify relevant questions for evaluating unfamiliar evidence using knowledge they already have. However, such translation must also flag when a method that looks unfamiliar is in fact standard practice in the secondary field (e.g., small-N interviews in ethnography), so that cross-domain translation calibrates critical thinking rather than manufacturing false suspicion.

\paragraph{Recommendation 2: Support cross-domain synthesis.}
While Recommendation 1 uses the researcher's primary domain to \textit{verify} claims in a secondary domain, interdisciplinary research also demands the reverse: \textit{combining} knowledge across domains to generate new insight. GenAI still struggles to make meaningful connections across multiple domains. It leaves researchers themselves to bridge disciplines, either through crafting prompts that use a particular structure to bridge multiple domains or by conducting manual synthesis after the fact. Our turn-level analysis also showed that knowledge synthesis rarely sustained itself across turns, unlike information-seeking, iterative refinement, and grounding. To fully support interdisciplinary research, future systems should function more as collaborative reasoning partners and not only producing information-dense outputs \cite{satyanarayan2024intelligence}. We envision \textbf{tools that offer two modes: (1) an information-seeking mode for broad exploration and (2) a knowledge-synthesis mode for integrating information already curated by the researcher for deeper analysis}. Such a design would help encourage an ideal co-research workflow for interdisciplinary researchers. This extends calls for mixed-initiative interfaces \cite{horvitz1999principles} and adaptive forms of control across different phases of human-AI interaction \cite{isaak2026}. Sensecape similarly demonstrates how interfaces can support movement between higher-level sensemaking and lower-level exploration \cite{suh2023sensecape}.

\paragraph{Recommendation 3: Support disciplinary-oriented output formats.}
As interdisciplinary researchers draw on multiple disciplines, they also encounter different conventions for representing and communicating knowledge. For instance, P1, working in mixed reality, expressed interest in richer multimedia outputs such as diagrams, slides, demo videos, and open-source code, while P13, working in American history and memory studies, valued maps and archival resources. \textbf{Adapting output formats to relevant disciplinary practices} could better integrate AI assistance into interdisciplinary workflows \cite{jiang2023graphologue}. This could help researchers move between fields while engaging with information in forms that reflect how knowledge is represented and used within each discipline. As multimodal GenAI systems continue to develop and mature, such format-adaptive support for interdisciplinary researchers can soon be within reach. 

\paragraph{Support prompting by translating the researcher's own questioning patterns.}

Controlling GenAI's response specificity can be challenging when crafting an effective prompt itself requires domain knowledge that interdisciplinary researchers may lack in their secondary domain. Participants (P9, P11, P12, P15) noted that careful prompting strategies can help shape answers, yet this assumes a level of expertise that novices may not have. Participants suggested that \textbf{the tool could go further by recommending relevant keywords}, highlighting preferred vocabulary for prompting, or offering guidance on how to frame effective prompts. P9 also encouraged having a community of interdisciplinary researchers exchange prompting strategies tailored to their domains: ``\textit{Experiences of other researchers would be really helpful [...] They can share, okay, you can use this specific prompt and wording to get the results that you want.}''

In addition to careful prompting strategies, GenAI could draw directly on how a researcher already asks questions in their primary field. Researchers likely have implicit patterns for how they pose questions in their primary domain, such as what they ask first, what follow-ups they use, what level of specificity they default to. GenAI could learn from these patterns and reformulate an analogous, well-formed question in the secondary domain, instead of leaving the researcher to reconstruct that same query structure from scratch in unfamiliar territory. It shifts prompting support from \textit{telling} researchers what vocabulary or framing to use, to \textit{translating} a questioning style they already have, into a new domain. This resonates with prior HCI work on scaffolding individual users' interactions with AI systems \cite{gajos2022people,subramonyam2025chi} and how communities can collaborate on prompt engineering in programming communities \cite{feng2024coprompt} and education content creation \cite{reza2025prompthive}.

\subsection{Limitations}
Our study has three main limitations. First, our sample of 15 interdisciplinary researchers provides rich qualitative evidence but limits generalizability across disciplines. Future work could examine these patterns with larger and more diverse samples. Second, the two-week study captures initial interactions and short-term orchestration strategies with Deep Research. Longer studies could examine how these practices change as researchers gain experience with research-oriented GenAI. Finally, participants used the free versions of OpenAI's Deep Research and ChatGPT, which at that time, restricted Deep Research to five uses.

Additionally, we note that the success ratings reflect participants' perceived helpfulness of GenAI, not verified correctness of GenAI outputs. This is a deliberate scope choice as we ask how researchers orchestrate GenAI based on their judgment, not whether that judgment was accurate.


\section{Conclusion}
Interdisciplinary researchers did not treat GenAI as an authoritative oracle, but mostly as a research assistant for information foraging. Our findings show that interdisciplinary GenAI use involves shifting epistemic positions: the same researcher moves between being an expert, a learner, and an integrator depending on the disciplinary context of their work. As these positions shift, so do researchers' needs from GenAI and their ability to evaluate its outputs. Irrelevant or imprecise outputs were sometimes tolerated when they provided useful directions or helped researchers move forward. Usefulness was therefore not defined by correctness alone, as researchers also valued inspiration, broader perspectives, and connections across fields.

These findings show how GenAI shapes the way researchers navigate disciplinary boundaries. While GenAI makes unfamiliar knowledge more accessible, researchers still need to evaluate and integrate that knowledge into their own research context while retaining epistemic ownership over their research. Future research assistants should support this process while preserving researchers' epistemic agency through multiple modes of engagement, flexible granularity of control, and designs that foreground human expertise rather than replace it.

\bibliographystyle{ACM-Reference-Format}
\bibliography{mybibliography}

\appendix
\section{Participant Interest Form}

\begin{figure}
    \centering
    \includegraphics[width=\linewidth]{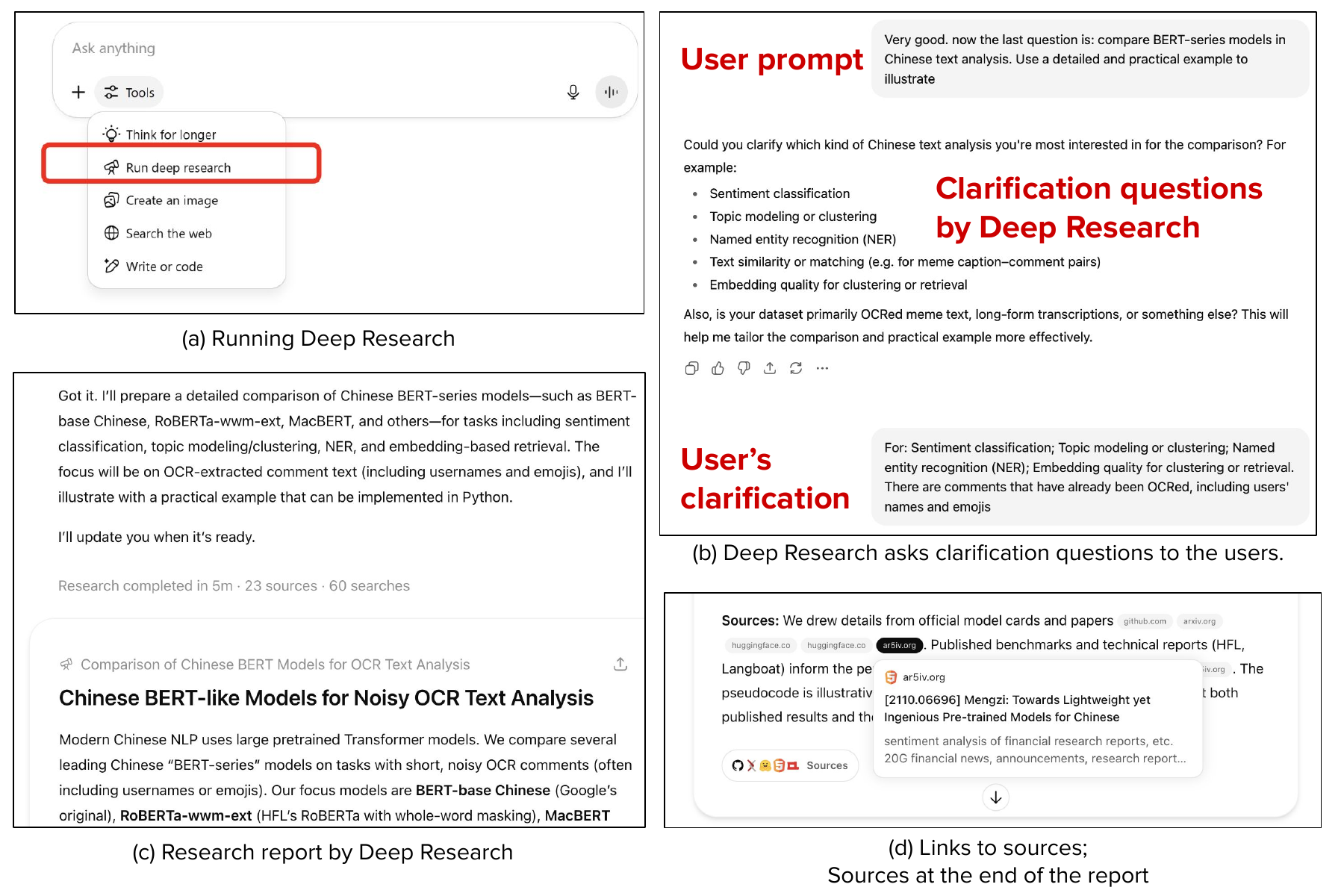}
    \caption{Deep research example}
    \label{fig:deep_research_example}
\end{figure}

\begin{enumerate}
    \item Email
    \item Full Name
    \item What is your major? 
    \item What is your research area?
    \item Are you currently leading or will you lead an interdisciplinary research project?
    \begin{itemize}
        \item I'm currently leading an interdisciplinary research project
        \item I will lead an interdisciplinary research project soon
        \item I'm currently leading a non-interdisciplinary research project
        \item I will lead a non-interdisciplinary research project soon
        \item I'm not leading any research project
    \end{itemize}
    \item If you are doing interdisciplinary research, what is your second research area?
    \item Are you fluent in English?
    \begin{itemize}
        \item Yes
        \item No
    \end{itemize}
    \item How many team members are in your project? (including your advisor)
    \item Are you willing to participate in our study if we collect your research progress?(e.g., writing, logging the research updates, etc.)
    \begin{itemize}
        \item Yes
        \item No
    \end{itemize}
\end{enumerate}

\section{Pre-interaction Survey}
\begin{enumerate}
    \item What is your major? 
    \item What is your first research area?
    \item I am confident with my knowledge for my research need in my first research area.(1 = Strongly disagree, 10 = Strongly agree)
    \item What is your second research area?
    \item I am confident with my knowledge for my research need in my second research area.(1 = Strongly disagree, 10 = Strongly agree) 
    \item What is your current research project title?
    \item Write a short description about your current research project.
    \item Which research stage are you at right now?
    \begin{itemize}
        \item Planning and ideation
        \item Literature review
        \item Data collection
        \item Data processing
        \item Experiment
        \item Analysis
        \item Writing manuscript
    \end{itemize}
    \item How familiar are you with generative AI tools as a user?(1 = Not at all familiar, 10 = Very familiar)
    \item How familiar are you with conducting research on large language models (LLMs) for technical purposes?(1 = Not at all familiar, 10 = Very familiar)
    \item How familiar are you with deep research tools as a user?(1 = Not at all familiar, 10 = Very familiar)
    \item If you have used deep research tools before, which tool(s) did you use?

    \subsection{Attitude toward Artificial Intelligence}
    \item I fear artificial intelligence.(1 = Strongly disagree, 10 = Strongly agree)
    \item I trust artificial intelligence.(1 = Strongly disagree, 10 = Strongly agree)
    \item Artificial intelligence will destroy humankind.(1 = Strongly disagree, 10 = Strongly agree)
    \item Artificial intelligence will benefit humankind.(1 = Strongly disagree, 10 = Strongly agree)
    \item Artificial intelligence will cause many job losses(1 = Strongly disagree, 10 = Strongly agree)

    \subsection{Other questions}
    \item What is your gender?
    \begin{itemize}
        \item Female
        \item Male
        \item Other
    \end{itemize}
    \item What is your age range?
    \begin{itemize}
        \item 18-25
        \item 26-35
        \item 36-45
        \item 46-55
        \item 56-65
        \item >65
    \end{itemize}
    \item What is the name of your institution (university, company)?
    \item What is your academic level?
    \begin{itemize}
        \item Undergraduate student
        \item Masters student
        \item Junior PhD student (year 1-3)
        \item Senior PhD student (year 4 or above)
        \item Postdoc
        \item Junior faculty
        \item Senior faculty
        \item Junior industry researcher
        \item Senior industry researcher
    \end{itemize}
\end{enumerate}

\section{Participant statistics}
\begin{figure}
    \centering
    \includegraphics[width=0.5\linewidth]{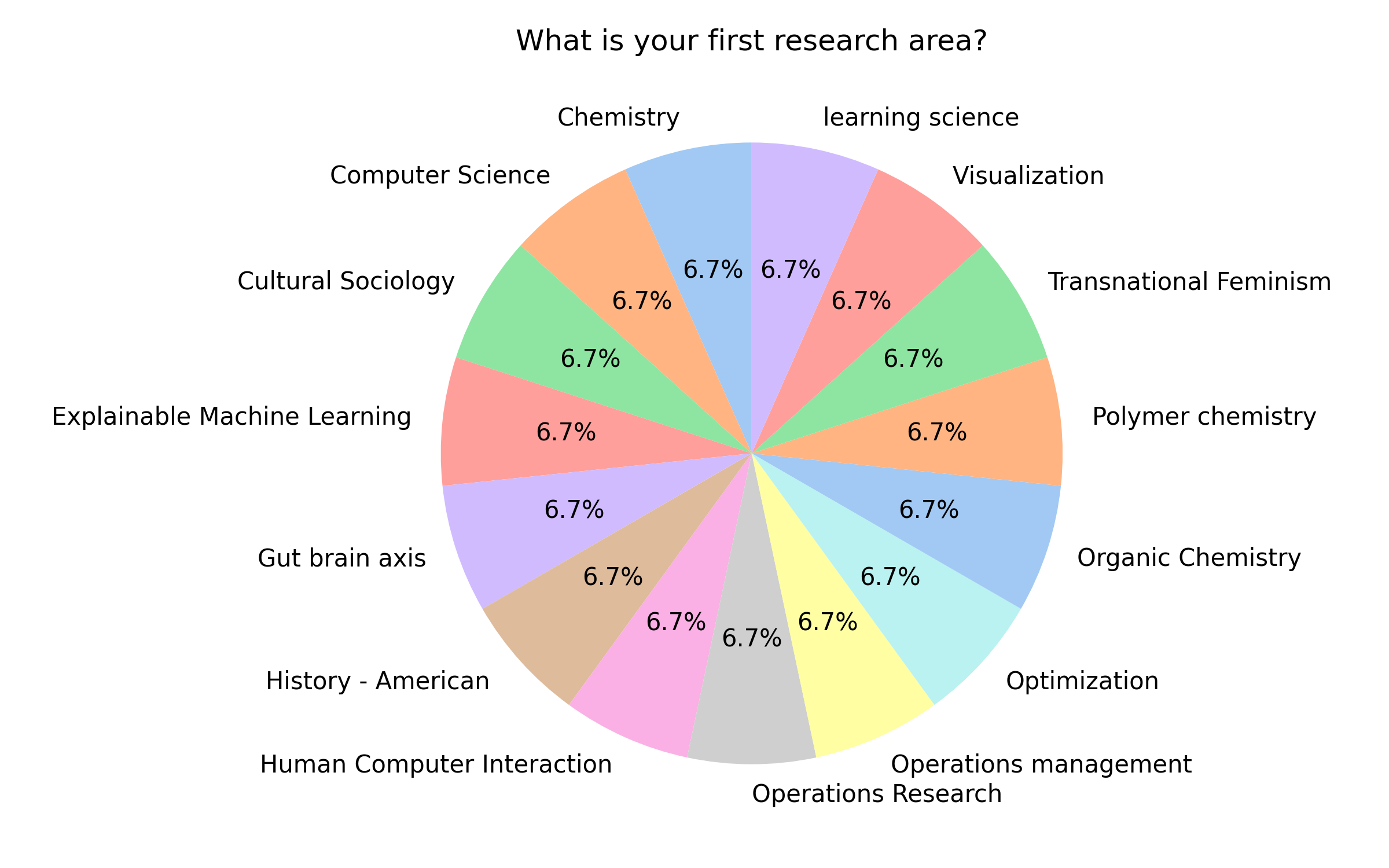}
    \caption{Responses to ``What is your first research area?``}
    \label{fig:What is your first research area}
\end{figure}

\begin{figure}
    \centering
    \includegraphics[width=0.5\linewidth]{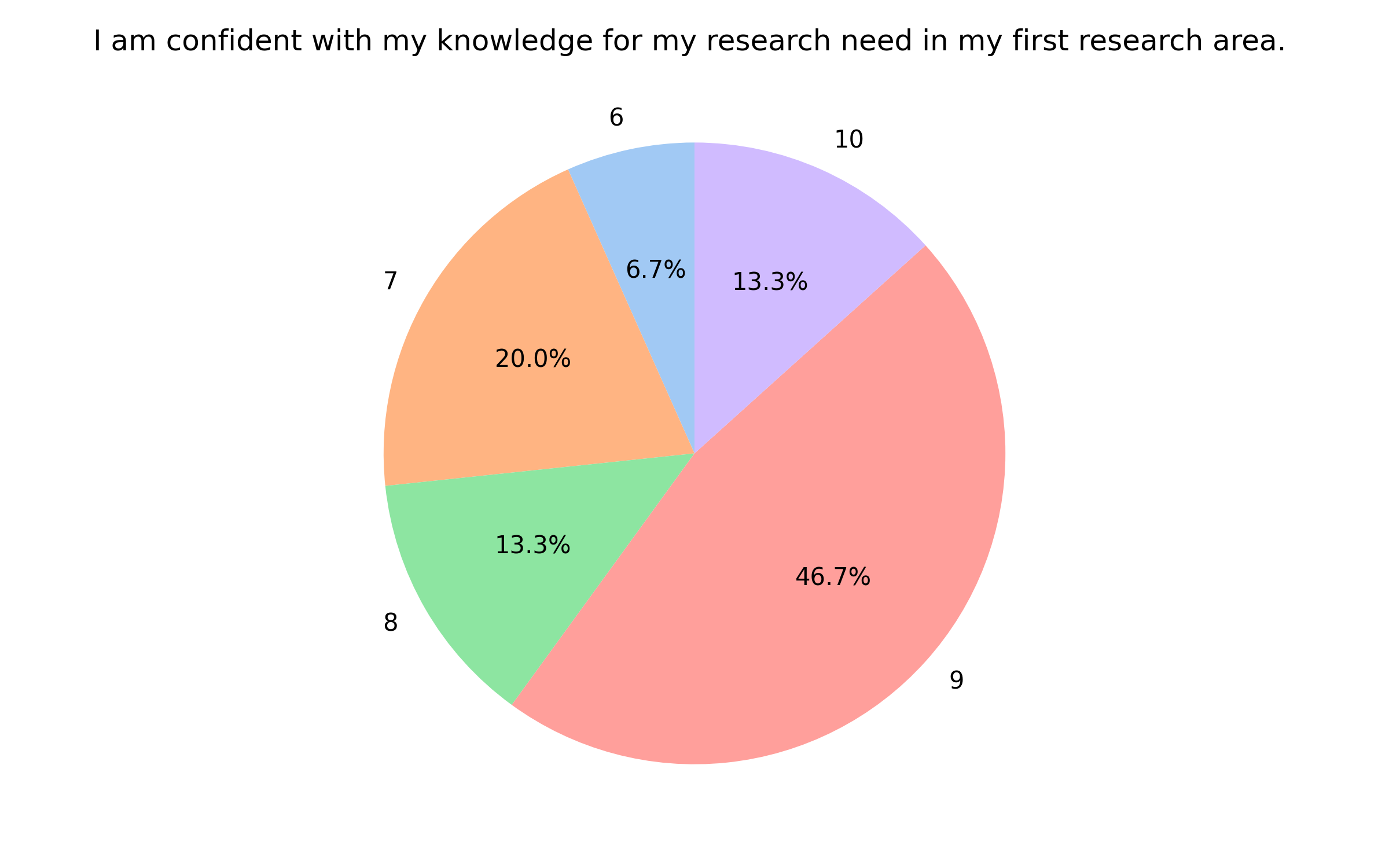}
    \caption{Responses to ``I am confident with my knowledge for my research need in my first research area.``}
    \label{fig:confident with my knowledge for my research need in my first research area.}
\end{figure}

\begin{figure}
    \centering
    \includegraphics[width=0.5\linewidth]{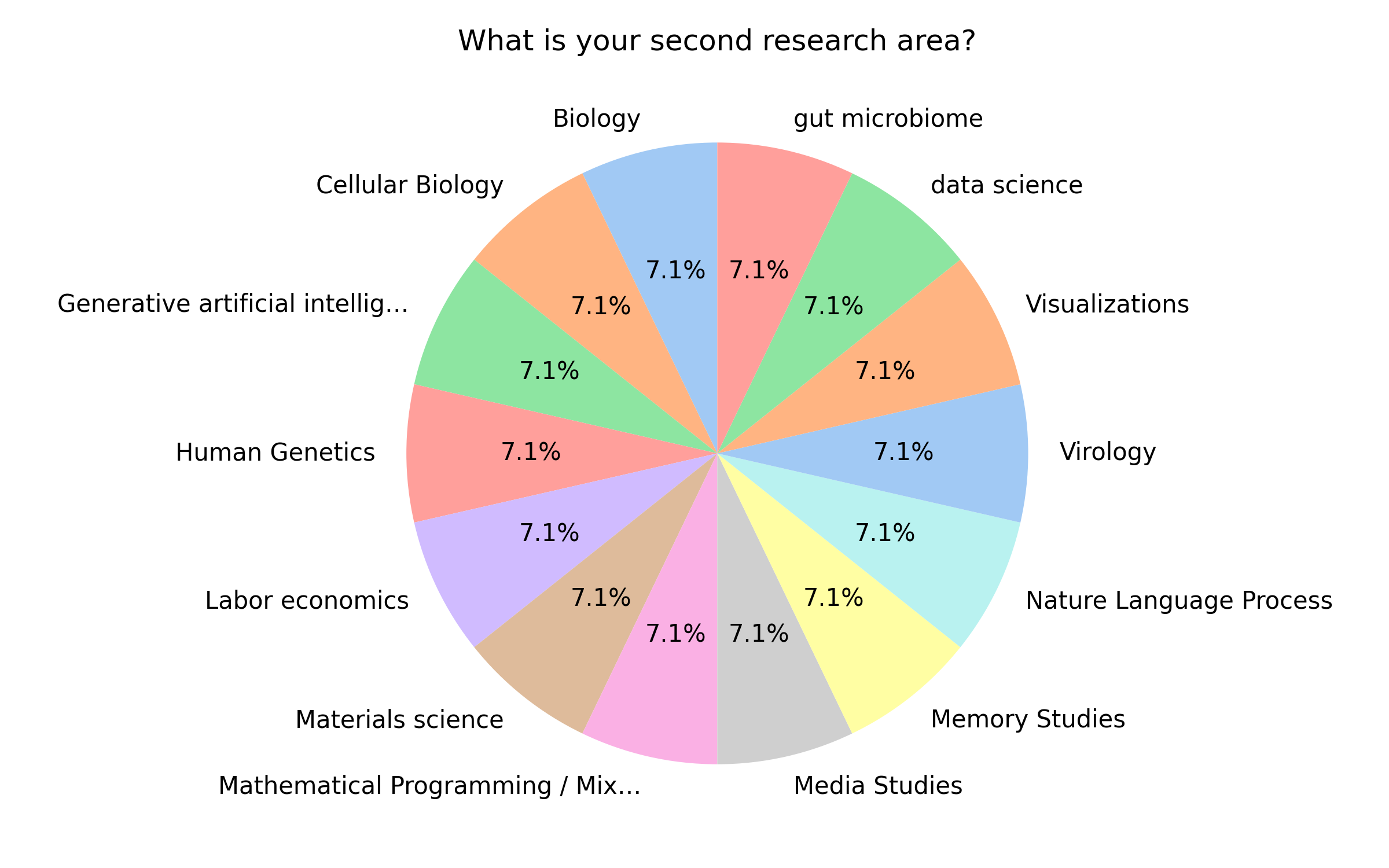}
    \caption{Responses to ``What is your second research area?``}
    \label{fig:What is your second research area}
\end{figure}

\begin{figure}
    \centering
    \includegraphics[width=0.5\linewidth]{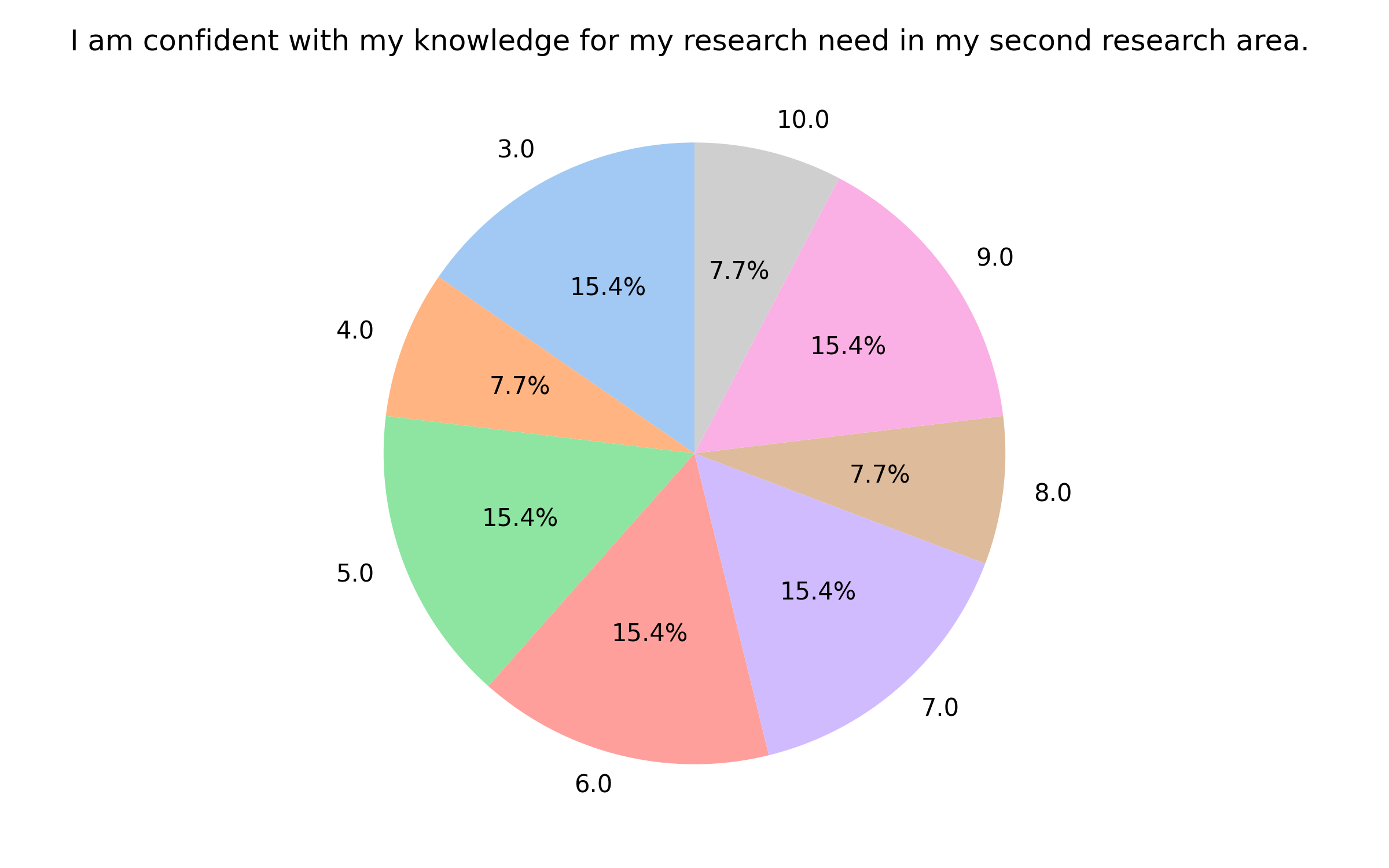}
    \caption{Responses to ``I am confident with my knowledge for my research need in my second research area.``}
    \label{fig:confident with my knowledge for my research need in my second research area.}
\end{figure}

Figure \ref{fig:What is your first research area}, Figure \ref{fig:confident with my knowledge for my research need in my first research area.}, Figure \ref{fig:What is your second research area}, Figure \ref{fig:confident with my knowledge for my research need in my second research area.} show participants' answers to the pre-interaction survey. 

\begin{table}[htbp]
\centering
\caption{Responses to ``Which research stage are you at right now?''}
\begin{tabular}{ll}
\hline
Participant & Research Stage \\
\hline
1 & Literature review \\
2 & Writing manuscript \\
3 & Data processing \\
4 & Planning and ideation \\
5 & Planning and ideation, Literature review, Data collection \\
6 & Planning and ideation, Literature review \\
7 & Writing manuscript \\
8 & Data collection, Experiment, Analysis \\
9 & Literature review \\
10 & Planning and ideation, Literature review, Data collection, Data processing \\
11 & Planning and ideation, Data collection \\
12 & I have received a major revision from a top-tier journal, and I am currently working on revising the paper. \\
13 & Writing manuscript \\
14 & Data collection, Data processing, Experiment, Analysis \\
15 & Data collection, Data processing, Experiment, Analysis, Writing manuscript \\
\hline
\end{tabular}
\end{table}

\begin{figure}
    \centering
    \includegraphics[width=0.5\linewidth]{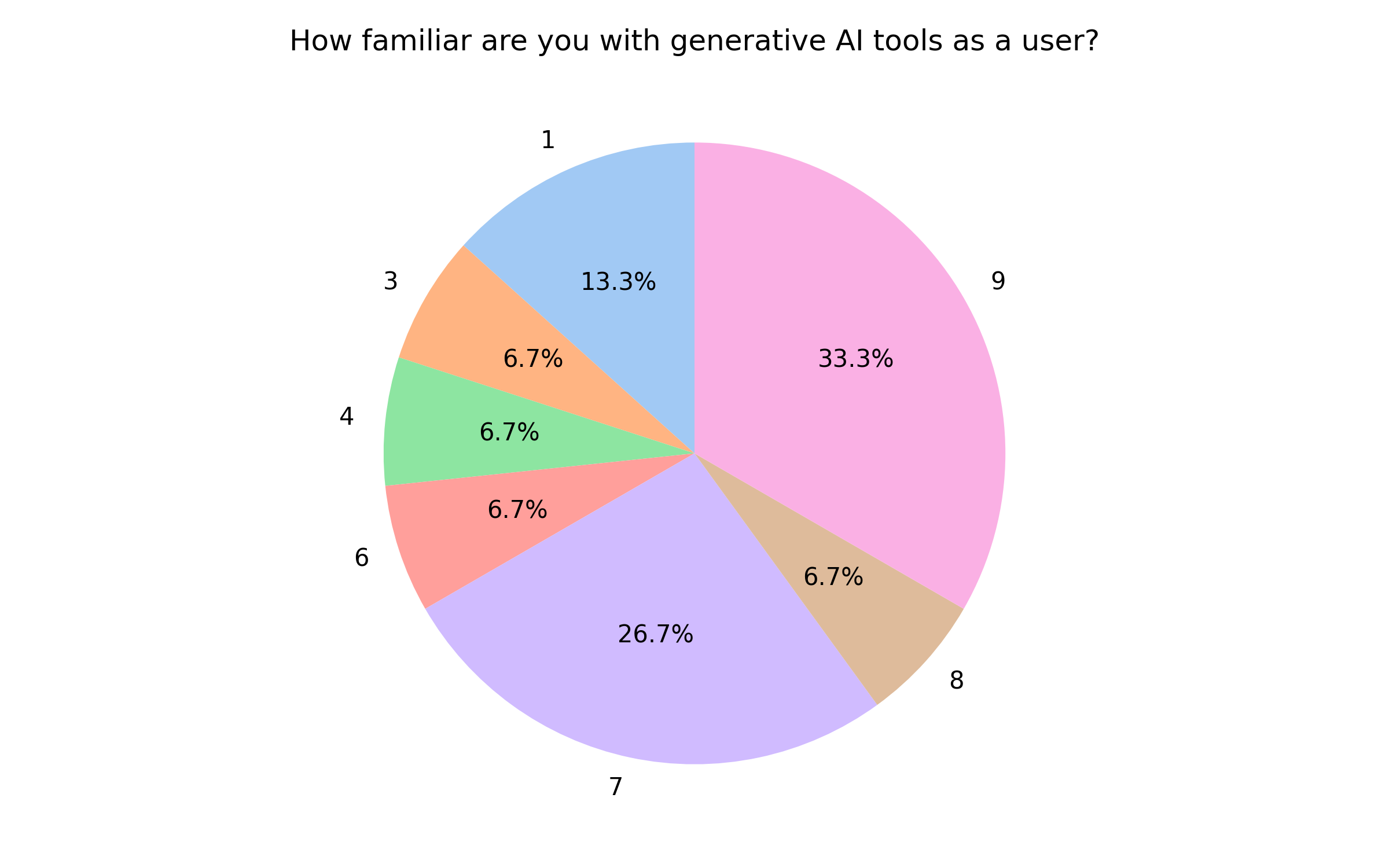}
    \caption{Responses to ``How familiar are you with generative AI tools as a user?``}
    \label{How familiar are you with generative AI tools as a user}
\end{figure}

\begin{figure}
    \centering
    \includegraphics[width=0.5\linewidth]{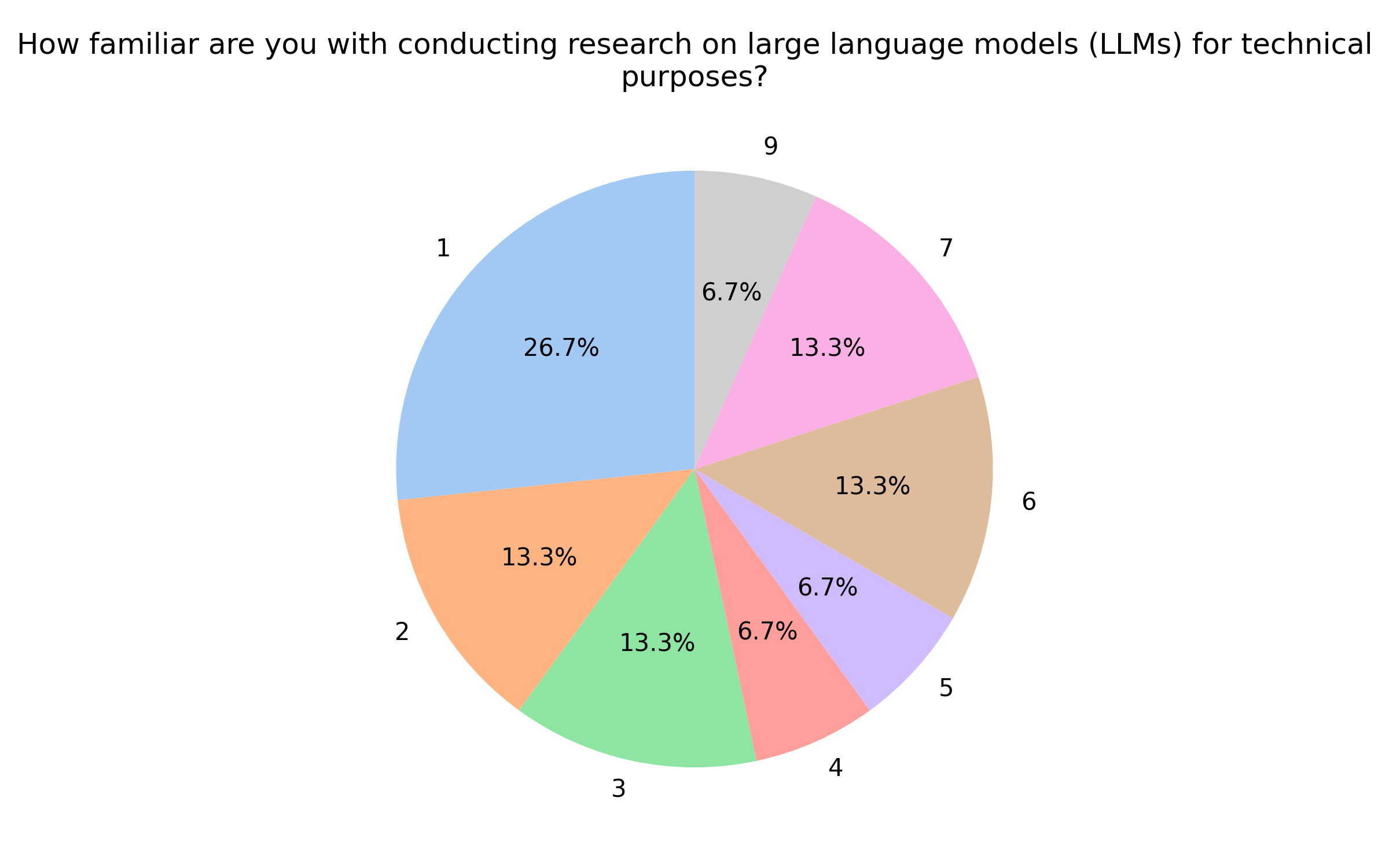}
    \caption{Responses to ``How familiar are you with conducting research on large language models LLMs for technical purposes?``}
    \label{How familiar are you with conducting research on large language models _LLMs_ for technical purposes}
\end{figure}

\begin{figure}
    \centering
    \includegraphics[width=0.5\linewidth]{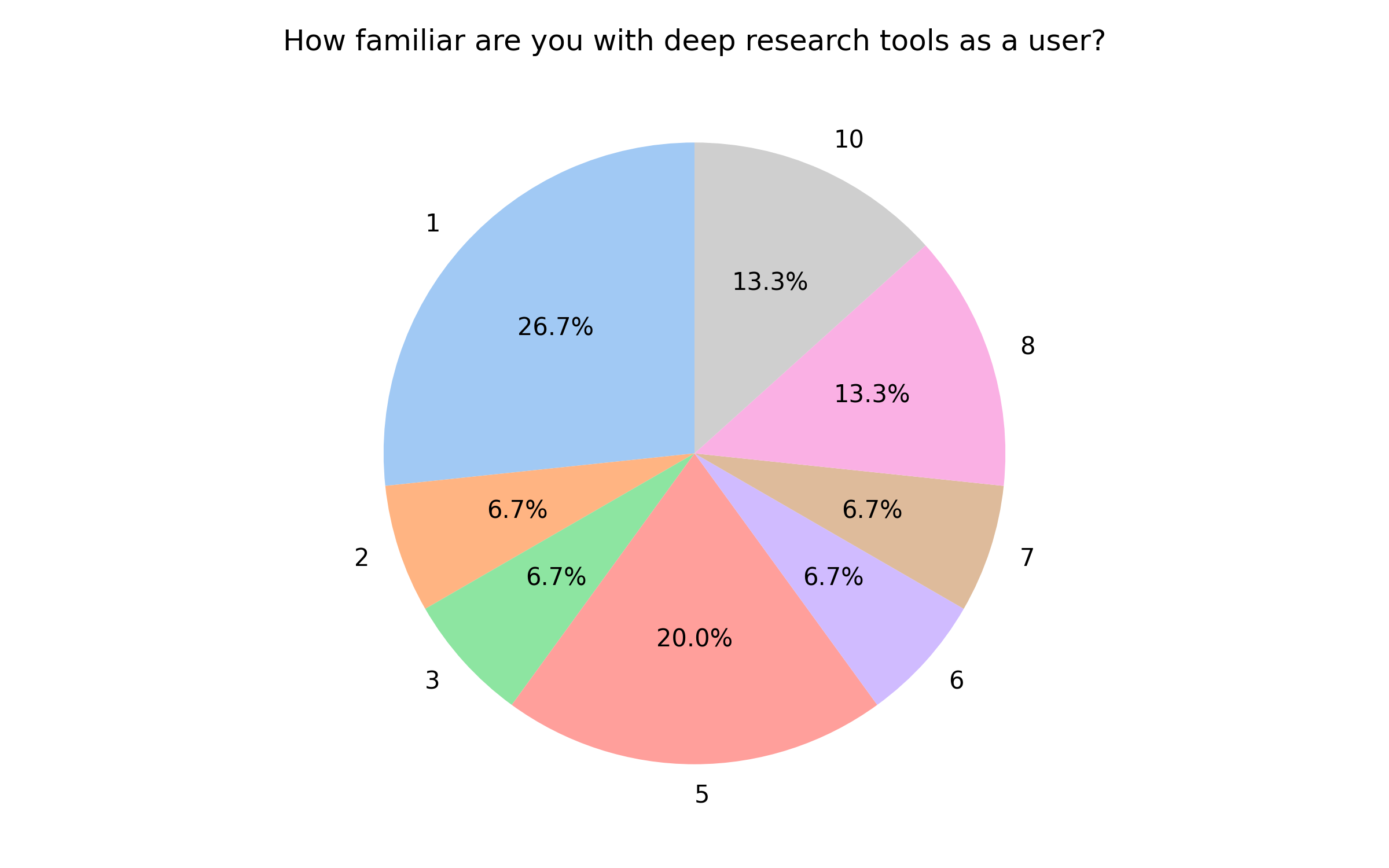}
    \caption{Responses to ``How familiar are you with deep research tools as a user?``}
    \label{How familiar are you with deep research tools as a user}
\end{figure}

\begin{table}[htbp]
\centering
\caption{Responses to ``If you have used deep research tools before, which tool(s) did you use?''}
\begin{tabular}{ll}
\hline
Participant & Tool(s) Used \\
\hline
1 & Mainly ChatGPT's deep research \\
2 & ChatGPT and deep seek \\
3 & ChatGPT's deep research \\
4 & Perplexity deep research \\
5 & ChatGPT and Perplexity \\
6 &  \\
7 & Gemini and ChatGPT's deep research tools \\
8 &  \\
9 & ChatGPT \\
10 & Gemini Deep Research \\
11 & ChatGPT's deep research \\
12 & ChatGPT's deep search + Claude research \\
13 &  \\
14 &  \\
15 &  \\
\hline
\end{tabular}
\end{table}

\begin{figure}
    \centering
    \includegraphics[width=0.5\linewidth]{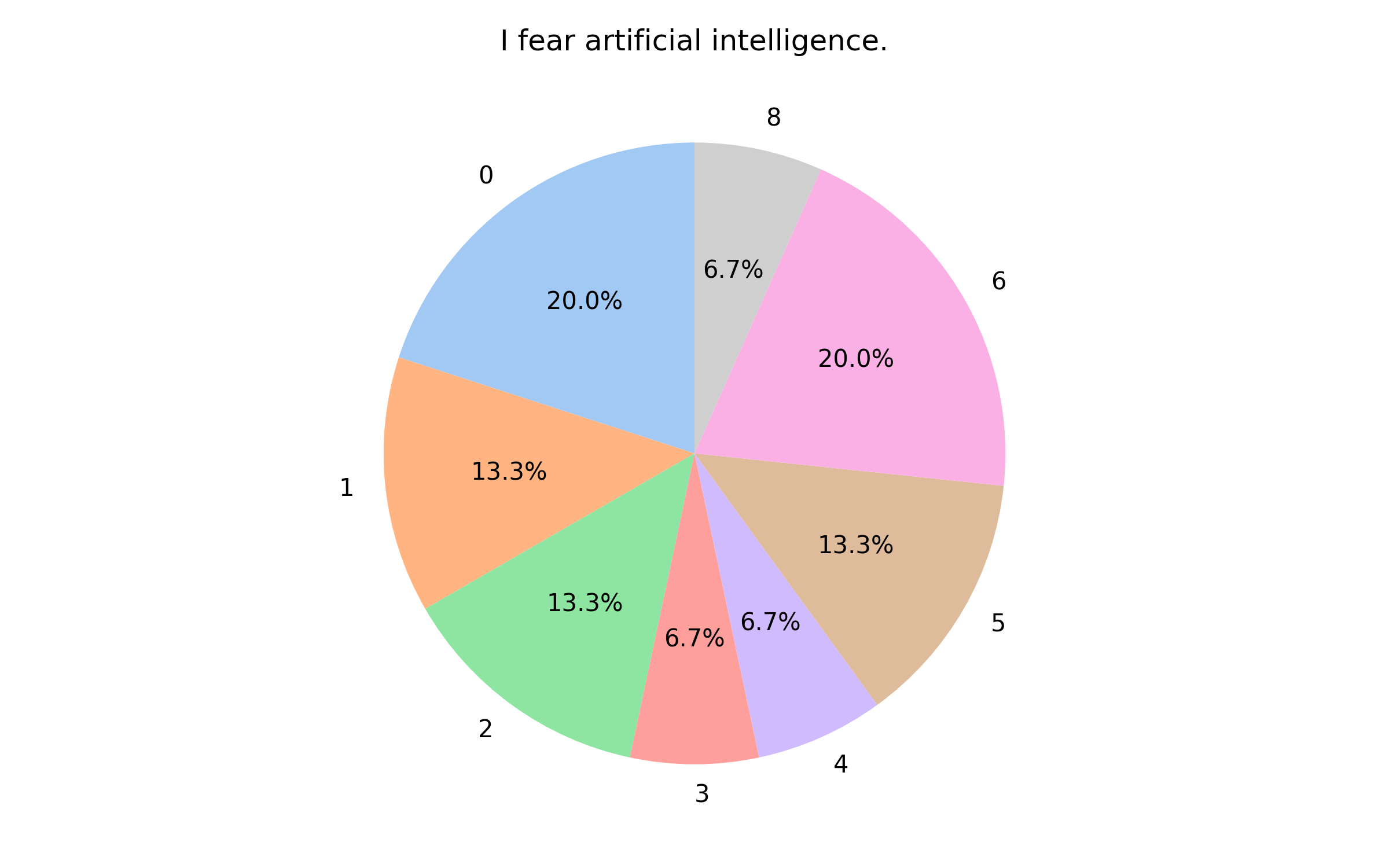}
    \caption{Responses to ``I fear artificial intelligence.``}
    \label{I fear artificial intelligence.}
\end{figure}

\begin{figure}
    \centering
    \includegraphics[width=0.5\linewidth]{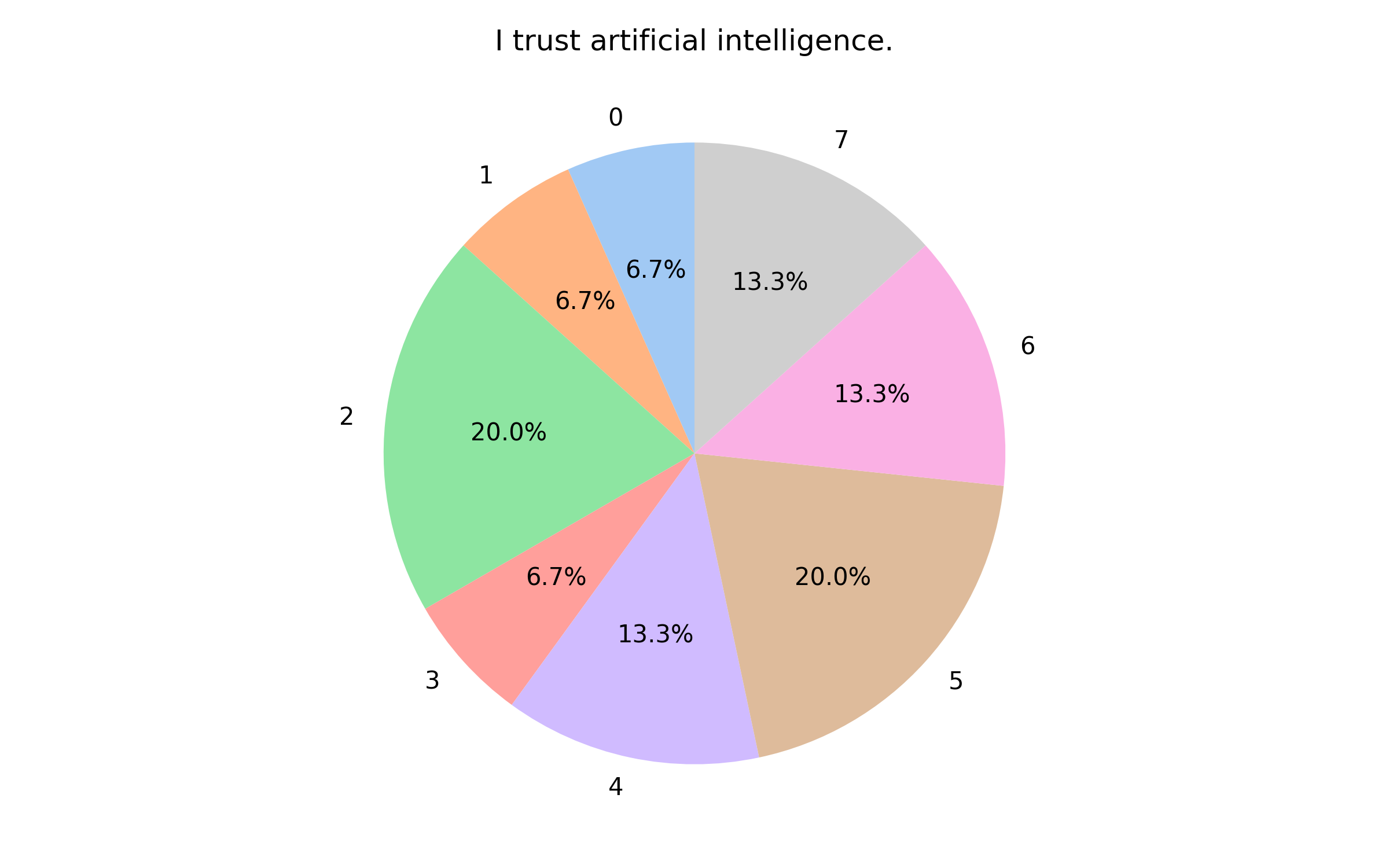}
    \caption{Responses to ``I trust artificial intelligence.``}
    \label{I trust artificial intelligence.}
\end{figure}

\begin{figure}
    \centering
    \includegraphics[width=0.5\linewidth]{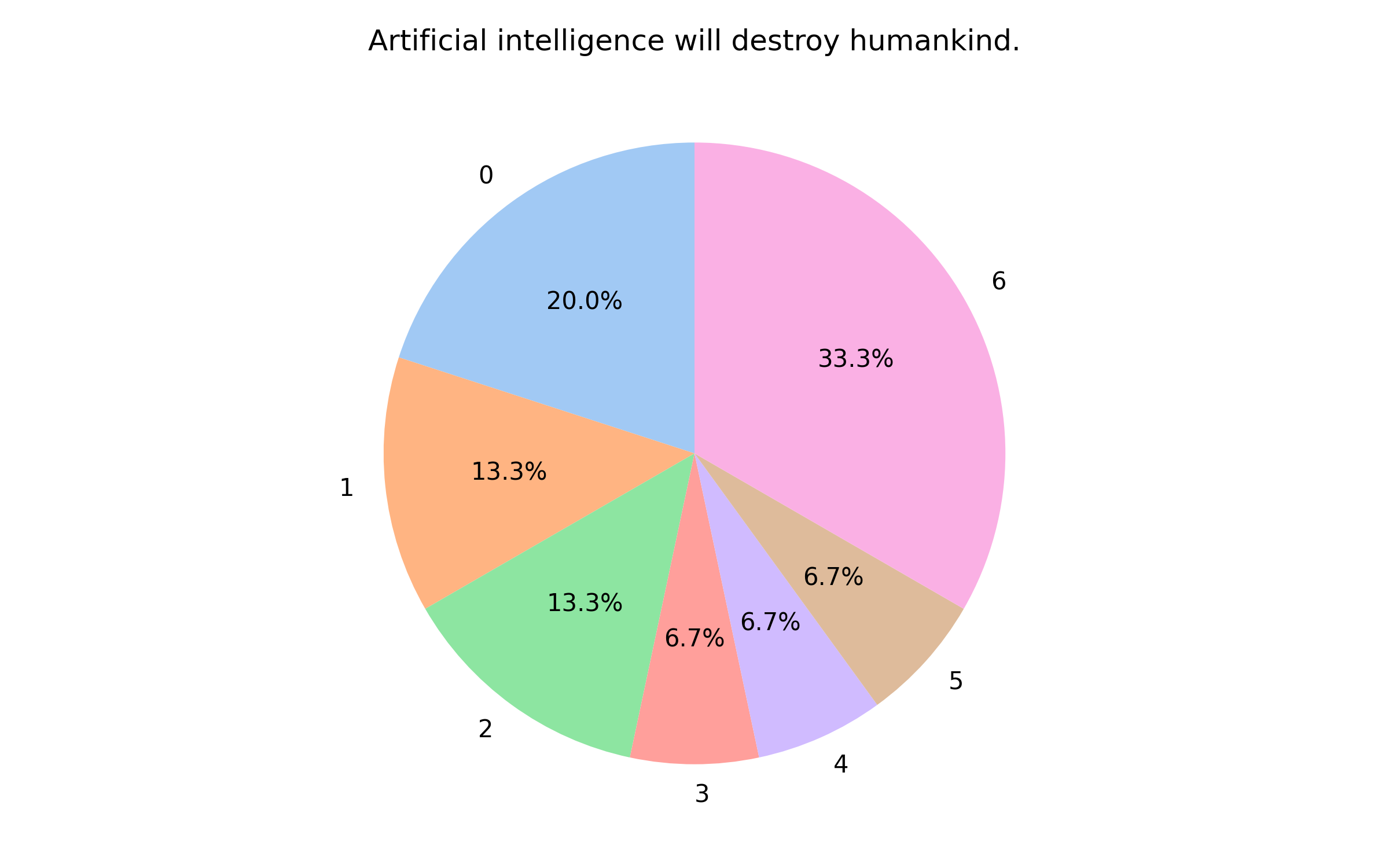}
    \caption{Responses to ``Artificial intelligence will destroy humankind.``}
    \label{Artificial intelligence will destroy humankind.}
\end{figure}

\begin{figure}
    \centering
    \includegraphics[width=0.5\linewidth]{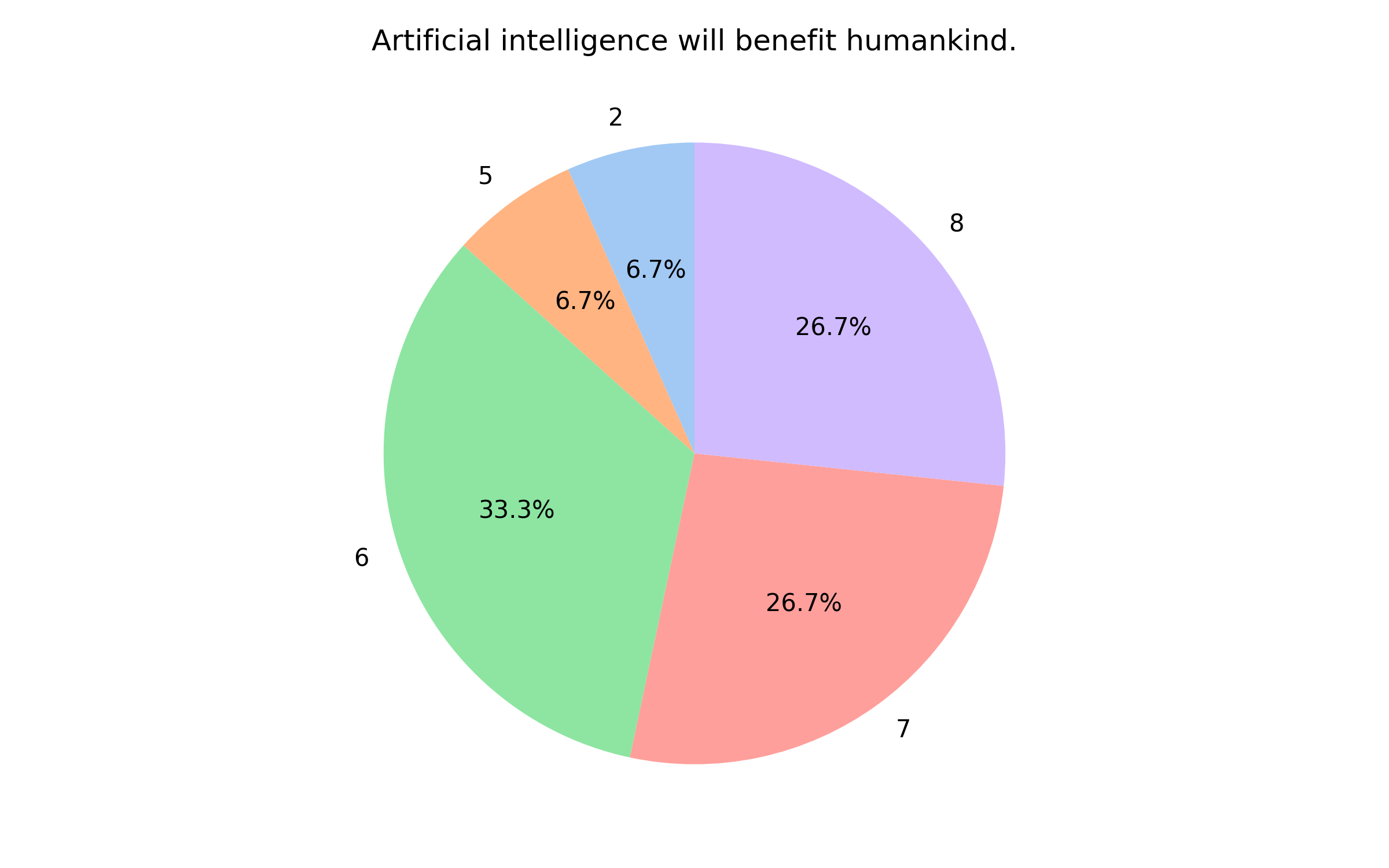}
    \caption{Responses to ``Artificial intelligence will benefit humankind.``}
    \label{Artificial intelligence will benefit humankind.}
\end{figure}

\begin{figure}
    \centering
    \includegraphics[width=0.5\linewidth]{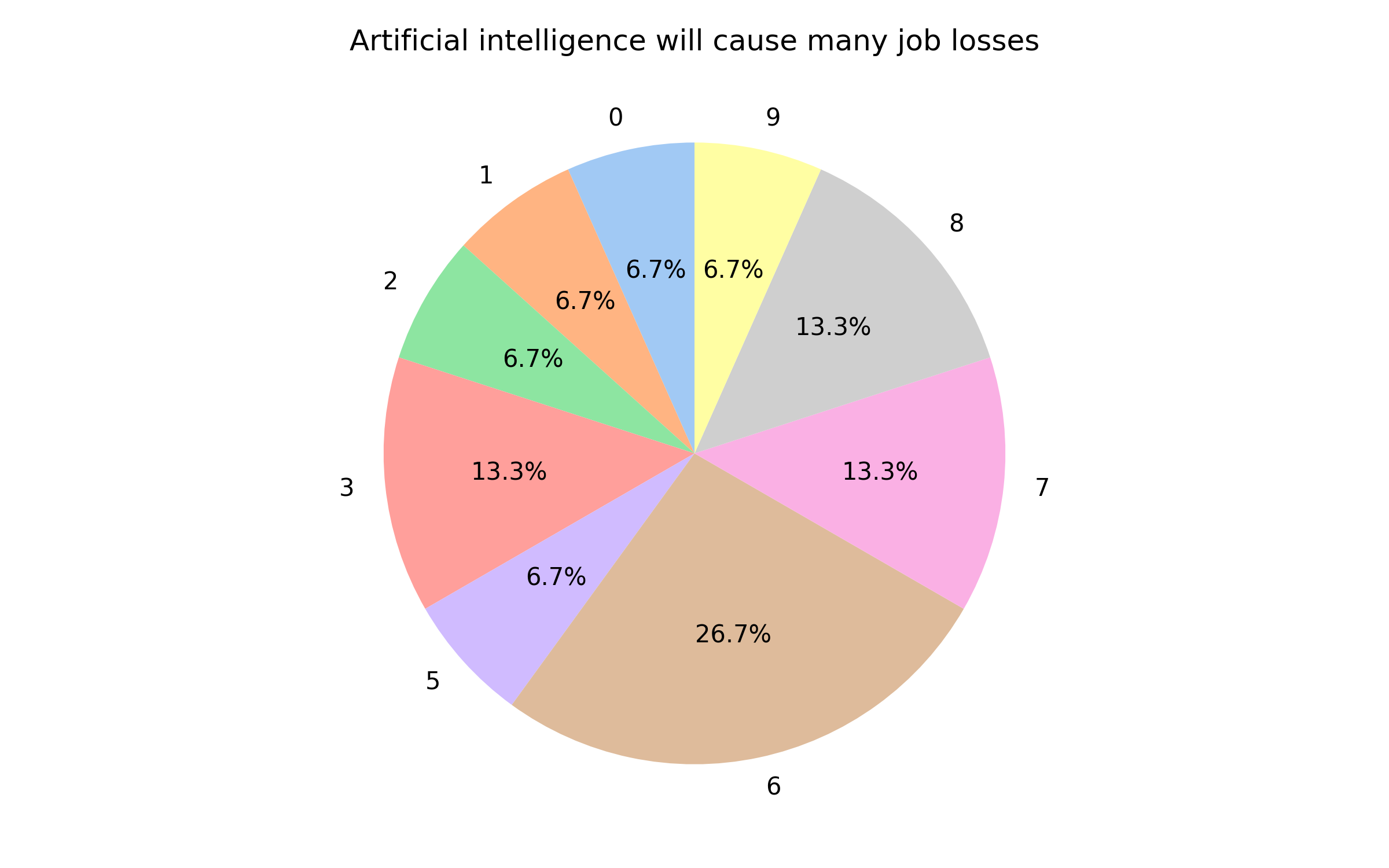}
    \caption{Responses to ``Artificial intelligence will cause many job losses.``}
    \label{Artificial intelligence will cause many job losses.}
\end{figure}

\begin{figure}
    \centering
    \includegraphics[width=0.5\linewidth]{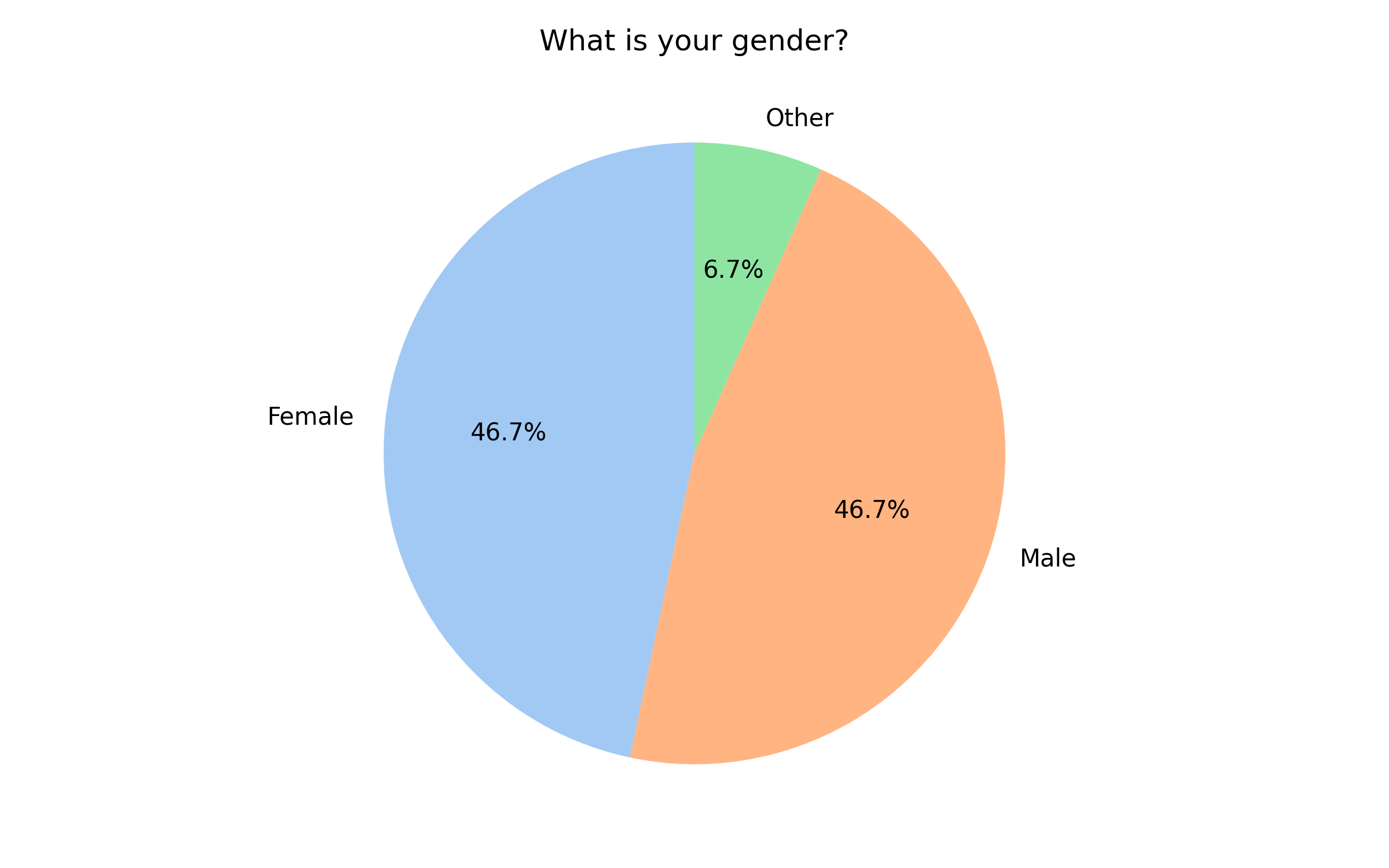}
    \caption{Responses to ``What is your gender?``}
    \label{What is your gender}
\end{figure}

\begin{figure}
    \centering
    \includegraphics[width=0.5\linewidth]{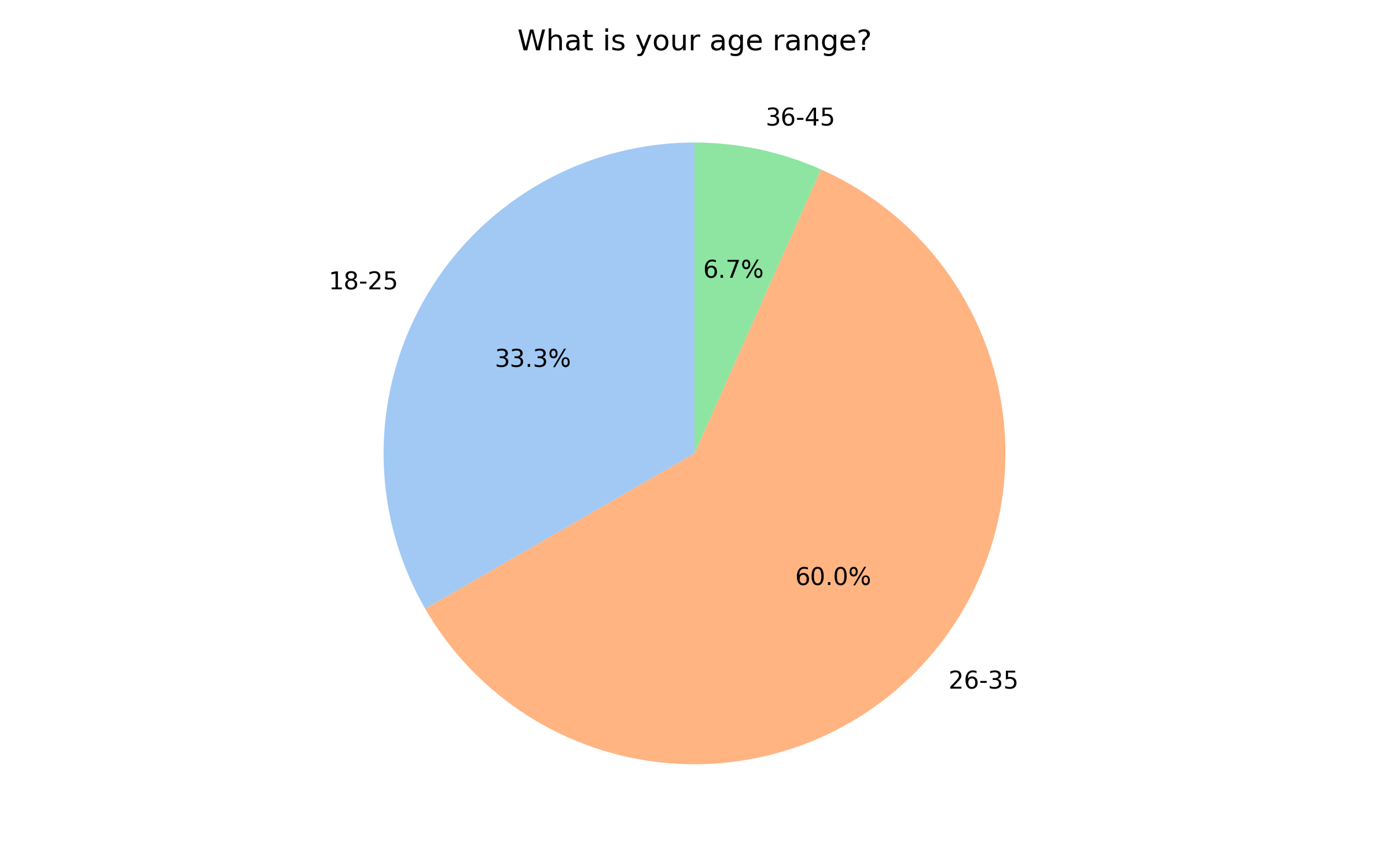}
    \caption{Responses to ``What is your age range?``}
    \label{What is your age range}
\end{figure}

\begin{figure}
    \centering
    \includegraphics[width=0.5\linewidth]{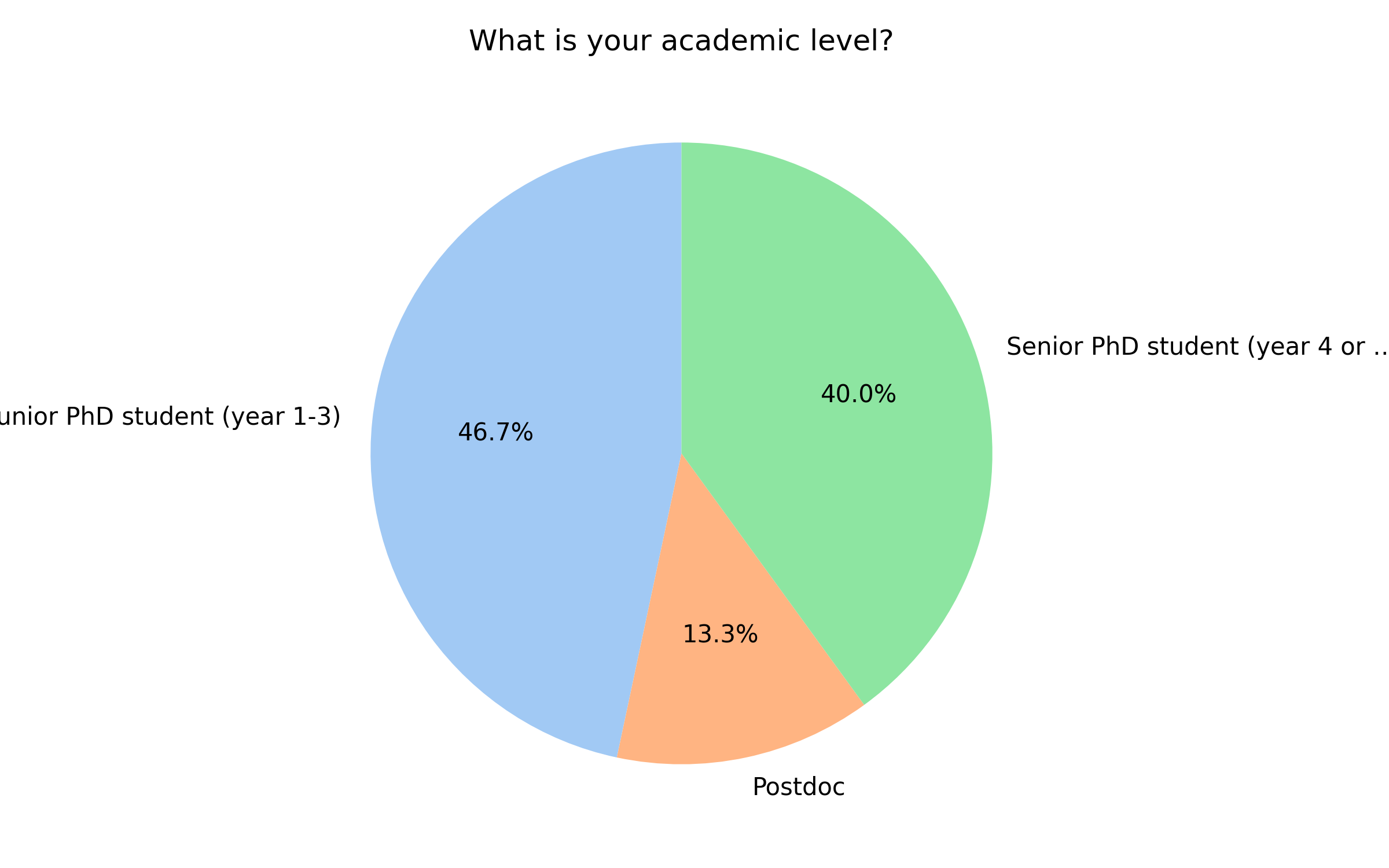}
    \caption{Responses to ``What is your academic level?``}
    \label{What is your academic level}
\end{figure}

\section{Prompt}
Table \ref{tab:prompt_label_part1} and Table \ref{tab:prompt_label_part2} show details about prompt labels, descriptions, and examples. 

\begin{table}[]
    \centering
    \begin{tabularx}{\textwidth}{ @{}p{4cm} | p{5cm} | p{5cm}@{} }
\toprule
\textbf{Label}    & \textbf{Description} & \textbf{Examples} \\ \midrule

\multicolumn{3}{@{}l}{\textbf{Grounding}} \\ \midrule
Declare background of research need & User provides background of research need. This usually happens at the initial prompt of a goal. & \textit{I would like to learn about literature-based discovery methods and workflows, as well as how computer science approaches have been used to automate or computationally implement these processes.} - P3 
\\ \hline
Provide information & User provides content/information about technical details. & \textit{error message: Error in longitudinal\_consensus\_cluster(data = disc\_data, id\_columns = ``person\_id'')} - P4 \newline 
\textit{the solvent I use was dry DCM and dry ACN. I added 5 drops of Net3} - P8
\\ \hline
Provide additional information & User provides additional information after LLM requests it (usually happens when running Deep Research) & \textit{The focus should be limited to human studies. I am looking for any evidence that can connect KDR to PD} - P11 \\ \midrule

\multicolumn{3}{@{}l}{\textbf{Information Seeking}} \\ \midrule
Ask for definition & User requests the meaning of a term or concept, usually starts with “What”. & \textit{what is use GAM} - P4 
\\ \hline
Ask for explanation & User asks LLM to obtain reasoning, mechanisms, or relationships behind a concept or phenomenon, usually for questions like “Why”, “How”, etc. & \textit{why maaslin3 yield fewer significant results than maaslin2} - P2 
\\ \hline
Ask for examples & User asks LLM to provide specific examples. & \textit{any example of using Metal for point cloud rendering} - P1
\\ \hline
Exploration & User requests generation of possibilities or alternative ways of approaching a task or problem based on existing information. & \textit{Search for papers that have flagged some critical issues in this direction.} - P5
\\ \hline
Brainstorming  & User prompts the LLM to generate multiple ideas or options without requiring depth or evaluation, encouraging creative and open-ended exploration. & \textit{Give me a suitable title for this work} - P5
\\ \hline
Technical question & User seeks domain-specific, procedural, or computational knowledge. These are typically “how-to” queries requiring methods, code, implementation steps, or applied statistical procedures, with an emphasis on actionable and precise outputs. & \textit{how to extract the content of the paper beyond abstract (e.g., introduction, methods) based on the link} - P3 \\ \hline

\multicolumn{3}{@{}l}{\textbf{Iterative Refinement}} \\ \midrule
Modify prompt & User revises their prompt with additional details to improve the LLM’s response. Unlike follow-up questions, this occurs when the initial answer is unsatisfactory. & \textit{The focus should be limited to human studies. I am looking for any evidence that can connect KDR to PD.} - P11 
\\ \hline
Follow-up question & User asks a new but related question that builds on the previous exchange. Unlike prompt modification, the original response is adequate; the user’s curiosity or understanding has simply progressed. & \textit{Using the same data, how can I estimate the maximum possible AUROC that can be achieved with this SNP heritability estimate?} - P11 
\\ \hline
Request clarification & User asks LLM to clarify or elaborate its previous response & \textit{what is taxa raw? I thought you called it taxa\_table} - P2 
\\ \hline
Positive feedback & User gives good feedback to LLM’s response & \textit{Yes, you are correct. And, you do not need to exclude any row.} - P8 
\\ \hline
Negative feedback & User gives bad feedback to LLM’s response, but no correction. & \textit{this code is not transposing} - P2 \\ 

\bottomrule
\end{tabularx}
\caption{Prompt label descriptions and examples (Part 1: Grounding, Information Seeking, Iterative Refinement).}
\label{tab:prompt_label_part1}
\end{table}

\begin{table}[]
    \centering
    \begin{tabularx}{\textwidth}{ @{}p{4cm} | p{5cm} | p{5cm}@{} }
\toprule
\textbf{Label}    & \textbf{Description} & \textbf{Examples} \\ \midrule

\multicolumn{3}{@{}l}{\textbf{Knowledge Synthesis}} \\ \midrule
Seeking validation & User seeks validation from LLM regarding user’s own information or writing. & \textit{is this okay for a manuscript?} - P2
\\ \hline
Ask LLM to analyze/process data & User asks the LLM to analyze or process data. & \textit{Analyze the results in this table} - P9
\\ \hline
Ask LLM to summarize & User asks the LLM to summarize paper or results. & \textit{Summarize your research} - P5
\\ \hline
Ask LLM to compare between two or more methods that the user already knew & User requests a comparison or integration of multiple pieces of information. & \textit{I want to run clustering with the temporal factor, like longitudinal clustering, to find out based on their behavioral patterns, and analyze how each cluster is related to different target variables, such as term GPAs, a frequency of specific SRL behaviors, and if they eventually stopped engaging or not. …. I specifically want you to focus on longitudinal clustering, its non-linear correlation/relationship analysis with target variables, and causal analysis.} - P4\\ \hline

\multicolumn{3}{@{}l}{\textbf{Miscellaneous}} \\ \midrule
Provide Scope & User writes constraints or boundaries to narrow the breadth of information or output style. This is to ensure the answer is aligned with the user’s needs. &   \textit{Papers do not have to be from education field. It could be from other fields like economics, biometrics, medical sciences, and so on.} - P4
\\ \hline
Acknowledgment & LLM often asks the user if they want to do something (run the code, create a table, etc.) and the user’s response to that LLM’s question is labeled as ``acknowledgement''.  & \textit{okay let's do that} - P2
\\ \hline
Others & Other prompts that don’t fall into all labels above go here, e.g., “discard previous context”.  & \textit{discard previous context} - P2\\

\bottomrule
\end{tabularx}
\caption{Prompt label descriptions and examples (Part 2: Knowledge Synthesis, Miscellaneous).}
\label{tab:prompt_label_part2}
\end{table}

\section{Goal labeling}
Labels for goal types are defined as follows. 
\begin{itemize}
    \item \textit{Technical information seeking}: Looking for specific technical related solutions which are not programming-related. Example goals: \textit{``Refine my methodology for metagenomics,'' ``Finding an alternative longitudinal data analysis methods''}. 
    \item \textit{Discover papers}: Searching for relevant academic literature. Example goals: \textit{``Collect papers using Apple ARkit and iPhone/iPad Lidar in HCI/Visualization/MR/Robotics``, '' ``Exploring literature on the effect of Medicaid expansion on health system capacity like number of hospitals or hospital beds; number of medical doctors or nurses``}. 
    \item \textit{Writing}: Seeking assistance with writing tasks, including revising their own writing or creating new content. Example goals: \textit{``Add subscript letters on my figures according to significance``, '' ``Write manuscript for ELISA Luminex methods``}.
    \item \textit{Programming}: Assistance with coding or technical problem-solving in programming. This label is different from \textit{technical information seeking} in a way that this label only deals with coding questions. Example goals: \textit{``I need to clear the errors I see when I adapt and run code``, '' ``Visualize multiple taxa together in ggplot (violin plots and boxplots)``}.
    \item \textit{Brainstorming}: Seeking creative responses or idea generation. Example goals: \textit{``Brainstorm ideas about how to form narratives for literature sensemaking``, '' ``Brainstorm ideas about how to form narratives for literature sensemaking``}.
    \item Summarization: Condensing information or extracting key points from larger text or papers. Example goals: \textit{``Summarize my draft and give critical suggestions``}.
    \item Suggestion: Asking for advice. Example goals: \textit{``I want to know more about cleaning data``, '' ``Summarize my draft and give critical suggestions``}.
    \item Proof: Verifying or validating information. Example goals: \textit{``Prove one theorem in the paper``}. 
    \item Data analysis: Asking for data analysis. Example goals: \textit{``Analyze GMM output using R``, '' ``Data analysis of experiment results on air-traffic management problems``}. 
\end{itemize}

\section{Prompt and goal label statistics}
\begin{table}[h]
\centering
\begin{tabular}{l|rr|rr|rr}
\toprule
\textbf{Label} 
& \multicolumn{2}{c}{\textbf{Deep Research}} 
& \multicolumn{2}{c}{\textbf{Standard GenAI}} 
& \multicolumn{2}{c}{\textbf{Both Tools}} \\
\midrule
Declare background of research need & 11 & 6.5\%  & 16 & 3.1\%  & 27 & 4.0\% \\
Provide information                 & 8  & 4.7\%  & 67 & 13.2\% & 75 & 11.1\% \\
Ask for definition                  & 2  & 1.2\%  & 13 & 2.6\%  & 15 & 2.2\% \\
Ask for explanation                 & 21 & 12.4\% & 45 & 8.9\%  & 66 & 9.7\% \\
Exploration                         & 17 & 10.0\% & 18 & 3.5\%  & 35 & 5.2\% \\
Brainstorming                       & 2  & 1.2\%  & 11 & 2.2\%  & 13 & 1.9\% \\
Ask for examples                    & 4  & 2.4\%  & 11 & 2.2\%  & 15 & 2.2\% \\
Technical question                  & 15 & 8.8\%  & 82 & 16.1\% & 97 & 14.3\% \\
Modify prompt                       & 8  & 4.7\%  & 27 & 5.3\%  & 35 & 5.2\% \\
Follow-up question                  & 3  & 1.8\%  & 54 & 10.6\% & 57 & 8.4\% \\
Request clarification               & 0  & 0.0\%  & 23 & 4.5\%  & 23 & 3.4\% \\
Seeking validation                  & 1  & 0.6\%  & 9  & 1.8\%  & 10 & 1.5\% \\
Ask LLM to analyze/process data     & 7  & 4.1\%  & 9  & 1.8\%  & 16 & 2.4\% \\
Ask LLM to summarize                & 2  & 1.2\%  & 8  & 1.6\%  & 10 & 1.5\% \\
Ask LLM to compare                  & 6  & 3.5\%  & 1  & 0.2\%  & 7  & 1.0\% \\
Positive feedback                   & 4  & 2.4\%  & 4  & 0.8\%  & 8  & 1.2\% \\
Negative feedback                   & 1  & 0.6\%  & 46 & 9.1\%  & 47 & 6.9\% \\
Provide scope                       & 9  & 5.3\%  & 6  & 1.2\%  & 15 & 2.2\% \\
Acknowledgment                      & 0  & 0.0\%  & 25 & 4.9\%  & 25 & 3.7\% \\
Others                              & 1  & 0.6\%  & 12 & 2.4\%  & 13 & 1.9\% \\
\midrule
\textbf{Total}                      & 170 & 100.0\% & 508 & 100.0\% & 678 & 100.0\% \\
\bottomrule
\end{tabular}
\caption{Prompt label frequency in Deep Research and Standard GenAI.}
\label{tab:prompt_stats}
\end{table}

Table \ref{tab:prompt_stats} shows prompt label frequency in Deep Research and Standard GenAI. Table \ref{tab:goal_prompt_stats} shows the number of goals and prompts that belong to which discipline (primary, secondary, or both). 

\begin{table}[h]
\centering
\begin{tabular}{lrrrr}
\toprule
 & Primary & Secondary & Both & Total \\
\midrule
Goals & 18 & 17 & 62 & 97 \\
Prompt-level coded instances & 75 & 147 & 456 & 678 \\
\bottomrule
\end{tabular}
\caption{Goal- and prompt-level statistics by disciplinary context.}
\label{tab:goal_prompt_stats}
\end{table}

\section{Normalization for Participant Imbalance}
\label{app:normalization}
Participants varied substantially in the number of conversational turns with GenAI, ranging from 6 to 94 turns across the 15 participants (411 total), and two participants alone accounted for 43.3\% of these turns. Because a single turn can carry multiple prompt-level codes, this imbalance is even more pronounced at the level of transitions: per-participant transition counts range from 14 to 348, and the same two participants account for 53.2\% of all 1182 transitions. A naive pooled count of transitions would therefore allow these high-volume participants to disproportionately shape the aggregate transition patterns, reflecting their individual conversational habits rather than patterns typical across the sample.

To mitigate this imbalance, we normalized transition frequencies within each participant before aggregating across participants. For each participant, we computed the proportion of their transitions corresponding to each state-to-state transition. State here is the prompt-level label. We then averaged these proportions across participants, assigning equal weight to each participant regardless of the number of turns they contributed. Formally, for a transition from state $a$ to state $b$, let $c_p(a,b)$ denote the raw count of that transition for participant $p$, and let
\begin{equation}
T_p = \sum_{a,b} c_p(a,b)
\end{equation}
denote the total number of transitions by participant $p$. We define the participant-normalized transition weight as
\begin{equation}
\hat{w}(a,b)
=
\frac{1}{N}
\sum_{p=1}^{N}
\frac{c_p(a,b)}{T_p},
\end{equation}
where $N=15$ is the number of participants.

Because each participant's transition proportions sum to one, the resulting normalized weights also sum to one:
\begin{equation}
\sum_{a,b} \hat{w}(a,b) = 1.
\end{equation}
For visualization, we subsequently rescale $\hat{w}(a,b)$ by the pooled raw transition total,
\begin{equation}
C = \sum_{a,b}\sum_{p=1}^{N} c_p(a,b),
\end{equation}
to obtain display weights
\begin{equation}
w_{\mathrm{display}}(a,b)
=
C\,\hat{w}(a,b).
\end{equation}
This rescaling is used only to keep edge widths visually comparable in magnitude to those in the raw transition diagram; it does not alter the relative proportions among transitions.

This normalization shifts weight away from transitions that are frequent primarily because of one or two high-volume participants and toward transitions that occur consistently across participants. For example, the \textit{Iterative-refinement} self-loop decreased from 104 transitions in the raw counts to 62.8 after normalization. This indicates that repeated refinement-on-refinement turns were disproportionately characteristic of the two high-volume participants rather than a pattern broadly shared across the sample.
\end{document}